\documentclass[12pt]{article}
\usepackage[utf8]{inputenc}
\pdfoutput=1 
\usepackage{amsfonts}
\usepackage{amsmath}
\usepackage{amssymb}
\usepackage{amsthm}

\usepackage[letterpaper, margin=3cm]{geometry}

\usepackage{graphicx} 
\usepackage[hang]{caption}

\usepackage[numbers, square,sort&compress,comma]{natbib}

\usepackage[bookmarks=false, colorlinks=true, linkcolor=blue, urlcolor=blue, citecolor=blue]{hyperref}

\usepackage{pifont}
\usepackage{setspace}

\newtheorem{co}{Corollary}
\newtheorem{lm}{Lemma}
\newtheorem{tr}{Theorem}

\newcommand*{ \jth}{$\mathrm{j} ^ {\text{th}}$ }
\begin{document}
\title{Perfectly matched layer for second-order time-domain elastic wave
equation: formulation and stability}

\author{Hisham Assi \footnote{\textit{Email address:} hisham.assi@mail.utoronto.ca}\quad,\quad Richard S. C. Cobbold\\ \small {Institute of Biomaterials and Biomedical Engineering, University of Toronto,}\\ \small{164 College Street, Toronto, M5S 3G9, Canada}}
\date{}
\maketitle


\begin{abstract}
A time domain system of equations is proposed to model elastic wave propagation in an unbounded two-dimensional anisotropic solid using perfectly matched layer (PML). Starting from a system of first-order frequency domain stress-velocity equations and using complex coordinate stretching approach with a two-parameter stretch function, a second-order formulation is obtained. The final system, which consists of just two second order equations along with four auxiliary equations, is smaller than existing formulations, thereby simplifying the problem and reducing the computational cost. The discrete stability of the solutions for a given mesh size is examined with the help of a plane-wave analysis of the corresponding continuous problem. It is shown that increasing the scaling parameter of the stretch function leads to significant stability improvements for certain anisotropic media that have known issues. Numerical computations for different isotropic and anisotropic media are used to illustrate the results. 

\end{abstract}

\tableofcontents
\doublespacing
\section{Introduction}
\label{sec:introduction}
\begin{figure}[!h]
\centering \includegraphics[width=0.9\textwidth]{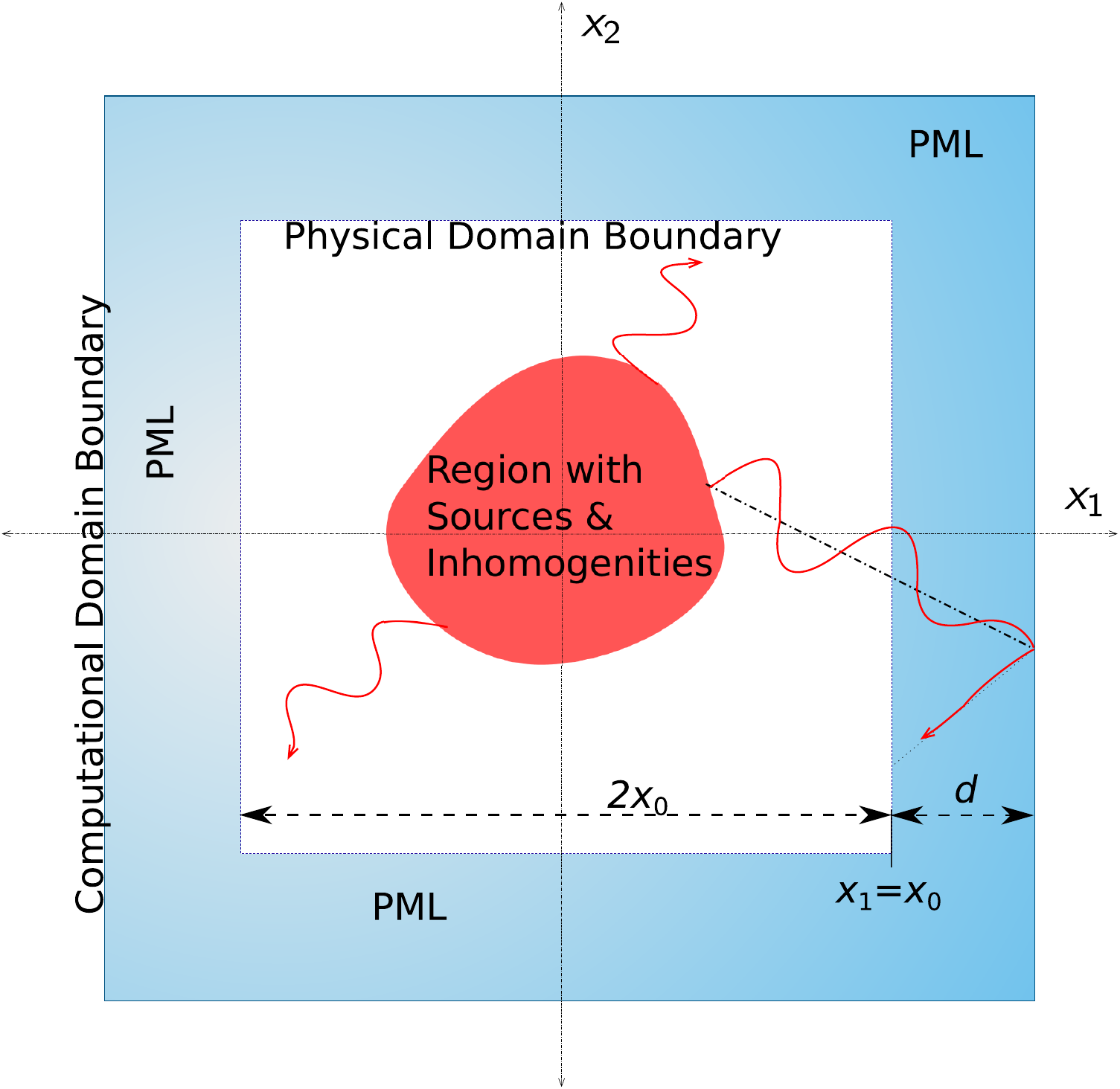} 
\caption{Illustrating the use of a perfectly matched layer (PML) for achieving near-perfect modeling of the solution to the unbounded wave radiation problem.}
\label{fig:illustrationPML} 
\end{figure}

Numerical simulations of wave propagation in an unbounded media  need special truncation methods  to avoid spurious wave reflections from the computational domain boundaries. Absorbing boundary conditions (ABCs) \cite{Engquist:1977tv} were first used. Such conditions work well when the waves are normally incident as in the case for 1D simulations, but this approach has limitations for higher dimensions. A more effective technique, as first described by Bérenger in 1994 \cite{Berenger:1994ua}, is to terminate the computational domain with a perfectly matched layer (PML). \autoref{fig:illustrationPML} illustrates the use of such a layer consisting of a hypothetical absorbing material that terminates the computational domain in such a way that the waves decay exponentially with negligible reflections from the outer boundaries, regardless of the incident angle. This is true for the case of an infinitely fine mesh i.e, for the continuous limit. In practice, a non-zero mesh element size causes some numerical reflections from the inner boundary of the PML, but these can be made very small, making PML an efficient means for modeling a variety of wave phenomena such as electromagnetic waves, acoustic waves in fluids, and elastic waves in solids.

For electromagnetic wave simulations, Bérenger \cite{Berenger:1994ua} showed that by adding specific conductivity parameters to Maxwell's equations perfect matching and decaying of the propagating waves in the PML could be achieved. An alternative method is to assume that the material contained within the PML is a uniaxial anisotropic media \cite{Sacks:1995gs,Roden:1997fo,Gedney:1996ub}, generally referred to as the uniaxial PML approach. In this method the original form of the wave equation is retained but with  frequency-dependent tensors as the material properties which makes it suitable for frequency domain simulations. A third method with greater generality and flexibility is the complex coordinate stretching approach \cite{Chew:1994dn}. In fact, the conductivity parameter introduced by Bérenger \cite{Berenger:1994ua} can be thought of as a parameter in a stretch function that extends the spatial coordinate in the layer to the complex plane. The addition of more parameters was subsequently proposed with the aim of making the method causal \cite{Kuzuoglu:1996kj}. Although the original PML was subsequently found to be causal \cite{Teixeira:1999bt,MezaFajardo:2008dx}, other benefits accrued from this new multi-parameter stretch function. Specifically, it was found that strong absorption occurred for the evanescent waves, improved absorption occurred at grazing angles \cite{Berenger:2002bq,Becache:2004fz,MezaFajardo:2008dx}, and improved stability was achieved in the PML for certain anisotropic elastic media \cite{Becache:2004fz,Duru:2012va,Appelo:2006wz}.

Many PML formulations have been introduced for elastic wave propagation \cite{Collino:2001vt,Hastings:1996um,Drossaert:2007fi,Chew:1996wk,Appelo:2006wz,MezaFajardo:2008dx,Kucukcoban:2011tn} as well as for general hyperbolic equations \cite{Appelo:2006vv}. Amongst these the split-field formulations usually make use of a single parameter stretch function and are typically described by systems of first order equations with double the number of physical equations such as those used by Bérenger \cite{Berenger:1994ua}. Unsplit field formulations use the physical fields variables along with extra auxiliary variables that are typically needed to obtain the time-domain equations from the frequency-domain equations. The use of multi-parameter stretch function usually requires a convolution to obtain a time-domain formulation, leading to the name convolutional PML \cite{Roden:2000cn} for many of the unsplit field models. The majority of these formulation uses a large number of equations (10 or more) to describe elastic wave propagation in the PML which affects the computational time and resources. Stability is  a known issue in PMLs \cite{Becache:2004fz,Duru:2012va, Appelo:2006wz, Loh:2009uc, Kreiss:2013tb, MezaFajardo:2008dx}, especially for  some  anisotropic solids. Some methods for  addressing this problem have been proposed \cite{Duru:2012va, Appelo:2006wz, MezaFajardo:2008dx}. In particular, by controlling  the stretch function parameters and the mesh size the discrete stability was improved for certain cases where the corresponding continuous problems were unstable \cite{Duru:2012va, Appelo:2006wz, Kreiss:2013tb}. 

The purpose of this paper is to introduce  second order time domain formulation for elastic wave propagation in isotropic and anisotropic solids in two space dimensions. Second order equations emerge directly from Newton's second law  which make them more robust as compared to the first order velocity-stress system of equations \cite{Kreiss:2013tb}. Moreover, the second order equations are more readily implemented in common numerical schemes \cite{Komatitsch:2007bz}, such as those used in PDE software packages like the finite element method-based (FEM) COMSOL Multiphysics (COMSOL, Inc., Burlington, Mass., U.S.A.) as used in this work. Other advantages accrue from using this formulation. First, it has a smaller number of equations than the classical and convolutional models, thereby simplifying the numerical implementation. Second, it has greater long-time stability for certain anisotropic media that are typically unstable in classical PML simulations. A simple method to further improve the discrete stability is proposed. In the next section we describe the background needed for obtaining the PML equations. This is followed by the derivation of our second order formulation. Then, with the help of a plane-wave analysis, the stability analysis is formulated. Numerical results are presented and discussed for both isotropic and anisotropic media. 

\section{Background and materials}
\label{sec:background}

\subsection{Elastic wave in solids }
\label{subsec:elasticWave}
The propagation of waves an elastic medium can be described using Newton's second law, Hook's law, and the linear approximation of the strain. These lead to the following three equations respectively: 
\begin{equation}
\rho\frac{\partial^{2}u_{i}}{\partial t^{2}}=\sum\limits _{j=1}^{d}\frac{\partial\sigma_{ij}}{\partial x_{j}},\label{eq:NewtonSolid}
\end{equation}
\begin{equation}
\sigma_{ij}=\sum\limits _{k,l=1}^{d}C_{ijkl}\,\varepsilon_{kl},\label{eq:HookSolid}
\end{equation}
\begin{equation}
\varepsilon_{kl}=\frac{1}{2}\left(\frac{\partial u_{k}}{\partial x_{l}}+\frac{\partial u_{l}}{\partial x_{k}}\right),\label{eq:strainLinear}
\end{equation}
where $u_{i}$ are the components of particle displacement vector, $\sigma_{ij}$, and $\varepsilon_{kl}$ are the components of the symmetric stress and strain tensors respectively, $C_{ijkl}$ are the components of the fourth order elasticity tensor with the following symmetries: $C_{ijkl}=C_{ijlk}=C_{jikl}$, and $C_{ijkl}=C_{klij}$, and $d$ is the number of space dimensions which is 2 for this work. The source of energy that excites the elastic medium can either be embedded in the boundary conditions or added as a load victor to \eqref{eq:NewtonSolid}. The above three equations together with the symmetry properties of the elasticity tensor enable the problem to be expressed as two second order equations in terms of the displacement vector: 
\begin{equation}
\rho\frac{\partial^{2}u_{i}}{\partial t^{2}}=\sum\limits _{j=1}^{2}\frac{\partial}{\partial x_{j}}\left(\sum\limits _{k,l=1}^{2}C_{ijkl}\frac{\partial u_{k}}{\partial x_{l}}\right).\label{eq:waveSoild}
\end{equation}

Another way to formulate the problem is through a system of first order equations in term of stress and velocity. These can be obtained using the same equations as used to obtain \eqref{eq:waveSoild}, leading to 
\begin{equation}
\begin{aligned}\rho\frac{\partial v_{i}}{\partial t} & =\sum\limits _{j=1}^{2}{\thinspace\frac{\partial\sigma_{ij}}{\partial x_{j}}}\\
\frac{\partial\sigma_{ij}}{\partial t} & =\sum\limits _{k,l=1}^{2}C_{ijkl\thinspace}\frac{\partial v_{k}}{\partial x_{l}}\,,
\end{aligned}
\label{eq:stressVelocity}
\end{equation}
where $v_{i}=\partial u_{i}/\partial t$ is the velocity vector component. In such a formulation five first order equations are needed to describe the problem. Namely, two velocity vector components, $v_{i}$ and four stress tensor components, $\sigma_{ij}$, which are reduced to three due to the symmetry in the stress tensor ($\sigma_{ij}=\sigma_{ji})$.

\subsection{Materials properties}
\label{subsec:materials}
All media considered in this work are orthotropic, which is a special case of an anisotropic media whose axes of symmetry coincide with $x_{1}$ and $x_{2}$. For such a medium the elasticity tensor has only four independent components. For simplicity and consistency with the notation commonly used \cite{Beltzer:1988ut}, we replace indices $11\to1$, $22\to2$, $12\to3$, and $21\to3$, so that the Hooks law for orthotropic media becomes 
\begin{equation}
\begin{pmatrix}\sigma_{1}\\
\sigma_{3}\\
\sigma_{3}\\
\sigma_{2}
\end{pmatrix}=\begin{pmatrix}C_{11} & 0 & 0 & C_{12}\\
0 & C_{33} & C_{33} & 0\\
0 & C_{33} & C_{33} & 0\\
C_{21} & 0 & 0 & C_{22}
\end{pmatrix}\begin{pmatrix}\frac{\partial u_{1}}{\partial x_{1}}\\
\frac{\partial u_{1}}{\partial x_{2}}\\
\frac{\partial u_{2}}{\partial x_{1}}\\
\frac{\partial u_{2}}{\partial x_{2}}
\end{pmatrix}.\label{eq:elasticityTensor}
\end{equation}
The elasticity coefficients are displayed in this notation in table \autoref{tab:material}. 
 
For the purpose of validation, testing, and stability analysis, we chose  five media whose characteristics are shown in \autoref{tab:material}. Material I is isotropic $(C_{11}=C_{22}=C_{33}+2C_{12}$) while the others are the anisotropic materials. In particular, media II, III, IV are identical to media II, III, IV as specified by Bécache \textit{et al.} \cite{Becache:2003ug}, and media V, which was also studied in \cite{MezaFajardo:2008dx, Duru:2012va},  corresponds to zinc crystal. 
The isotropic medium was used to test our PML and, by comparison with theoretical predictions, to validate the results of our numerical simulations. The anisotropic media was mainly used to study the stability.
\begin{table}[h!]
\caption{Elasticity coefficients for the materials examined.}
\centering %
\begin{tabular}{lcccc}
\hline
\hline 
Material  & $C_{11}$  & $C_{22}$  & $C_{33}$  & $C_{12}$ \tabularnewline
\hline 
I & 7.8 & 7.8 & 2  & 3.8 \tabularnewline
 II  & 20  & 20  & 2  & 3.8 \tabularnewline
III  & 4  & 20  & 2 & 7.5 \tabularnewline
 IV  & 10  & 20  & 6  & 2.5 \tabularnewline
 V  & 16.5 & 6.2 & 3.96 & 5\tabularnewline
\hline
\hline 
\end{tabular}\label{tab:material}	
\end{table}

\subsection{Plane waves and slowness curves}
\label{subset:planeWave}
To better understanding the wave propagation properties for equation like \eqref{eq:waveSoild}, it is useful to consider plane wave solutions of the form
\begin{equation}
\mathbf{u}=\mathbf{u}_{0}e^{i(\mathbf{k\cdot x}-\omega t)},\label{eq:harmonicWave}
\end{equation}
where $\mathrm{\mathbf{u}}_{\mathrm{\mathbf{0}}} \in \mathbb{C}^2$ is the polarization vector, or the amplitude of the wave with wavevector $\mathrm{\mathbf{k}} \in \mathbb{R}^2$ and angular frequency $\omega \in \mathbb{C}$, and $i=\sqrt{-1}$. The dispersion relation between $\mathrm{\mathbf{k}}$ and $\omega$, can be obtained by substituting \eqref{eq:harmonicWave} into \eqref{eq:waveSoild}. Assuming $\rho=1$ and that $C_{ijkl}$ are constants this results in a fourth order polynomial given by 
\begin{equation}
F_{0}\left(\omega,\thinspace\mathrm{\mathbf{k}}\right)=\det{\left(\omega^{2}\delta_{ik}-\sum\limits _{j,l=1}^{2}{C_{ijkl}k_{j}k_{l}}\right)=0},\thinspace\label{eq:dispersionElastic}
\end{equation}
where $\delta_{ik}$ is the Kronecker delta function. For an orthotropic medium this can be written as 
\begin{equation}
\begin{aligned}F_{0}\left(\omega,k_{1},k_{2}\right) & =\omega^{4}-\omega^{2}\left[\left(C_{11}+C_{33}\right)k_{1}^{2}+\left(C_{33}+C_{22}\right)k_{2}^{2}\right]+C_{11}C_{33}k_{1}^{4}\\
 & \quad+C_{22}C_{33}k_{2}^{4}+\left(C_{11}C_{22}-c_{12}^{2}-2C_{12}C_{33}\right)k_{1}^{2}k_{2}^{2}=0\thinspace,
\end{aligned}
\label{eq:dispersionOrthotropic}
\end{equation}
which is the characteristic polynomial of \eqref{eq:waveSoild} for the orthotropic case. We will refer for the four roots of \eqref{eq:dispersionOrthotropic}, $\omega_{n}\left(k_{1},k_{2}\right)$ where $\left(n=1...4\right)$ as the physical modes. 

Consider the following two conditions on the elasticity tensor 
\begin{equation}
\begin{aligned} & C_{11}>0,\, C_{22}>0,\, C_{33}>0,\,\text{and }C_{11}C_{22}>C_{12}^{2}\\
 & C_{11}\ne C_{33}\text{and }C_{22}\ne C_{33}.
\end{aligned}
\label{eq:ellipticityConditions}
\end{equation}
If the first condition is satisfied then the four roots of \eqref{eq:ellipticityConditions} are all real. Moreover, if the second condition is also satisfied then the four roots will be distinct enabling the group velocity to be defined by 
\begin{equation}
\mathbf{V}_{g}=\nabla_{\mathbf{k}}\omega=-\frac{\nabla_{\mathbf{k}}F_{0}\left(\omega,\mathbf{k}\right)}{\partial F_{0}\left(\omega,\mathbf{k}\right)/\partial\omega}\label{eq:groupVelocity}
\end{equation}
which specifies the direction of energy transport. The slowness vector defined by $\mathbf{S}=\mathbf{k}/\omega$ provides a convenient means for understanding the dispersion relations. Since \eqref{eq:ellipticityConditions} is homogeneous in $\mathbf{k}$ and $\omega$, it can be expressed as 
\begin{equation}
F_{0}\left(1,\thinspace S_{1},S_{2}\right)=0.\label{eq:SlowVelocity}
\end{equation}

\begin{figure}[!h]
\centering \includegraphics[width=1\textwidth]{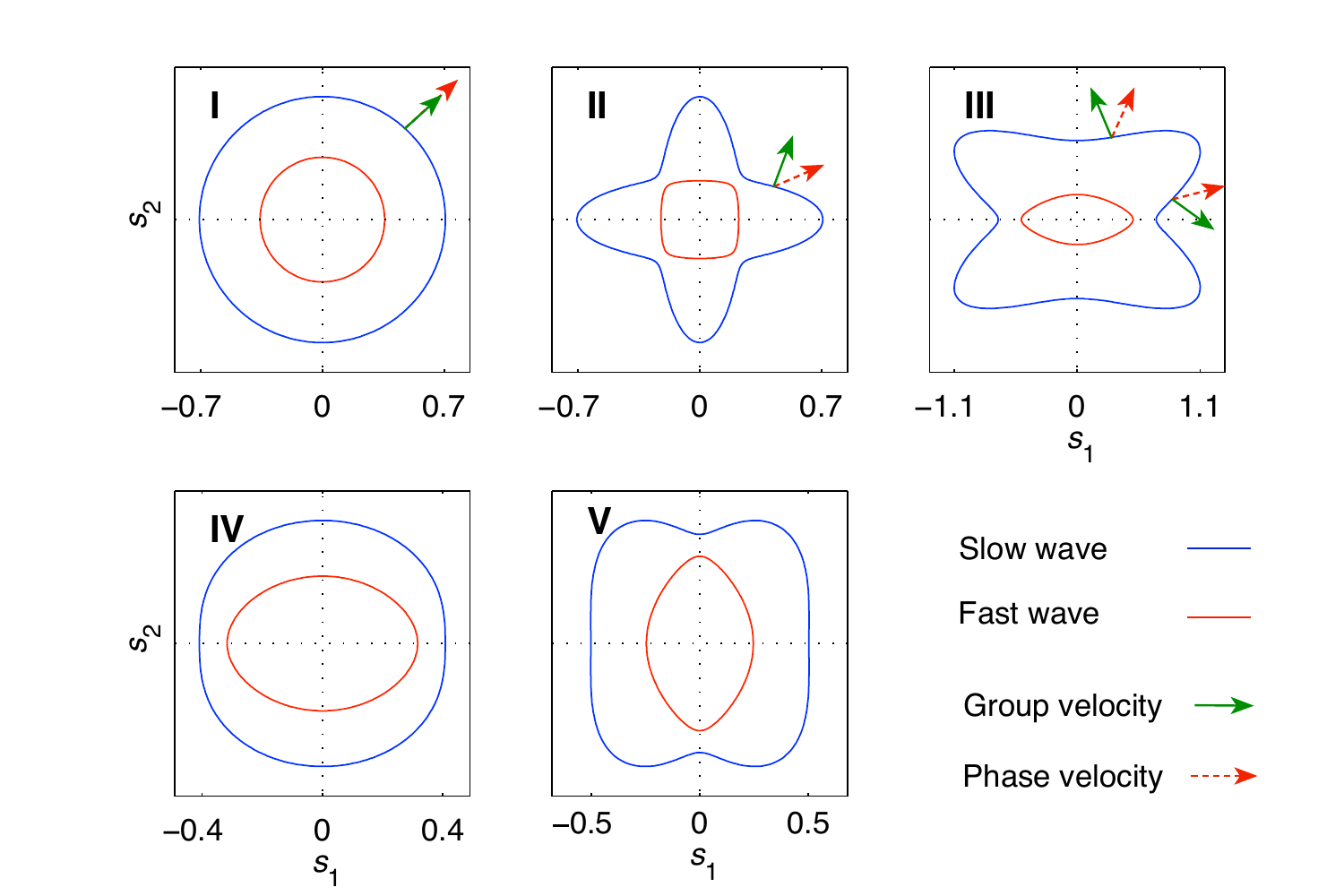} 
\caption{Slowness curves for all the materials whose properties are given in \autoref{tab:material}. Phase and group velocity are indicated for selected points.}
\label{fig:slowness} 
\end{figure}

For the materials in \autoref{tab:material} slowness curves, which are the plot of \eqref{eq:SlowVelocity}, are shown in \autoref{fig:slowness}. The inner curve corresponds to the fast wave (the longitudinal or quasi-longitudinal) and the outer curve corresponds to slow waves (shear or quasi-shear). The phase velocity, $V=\omega/\left|\mathbf{k}\right|=\pm1/\left|\mathbf{S}\right|$, in each propagation direction can be obtained from the slowness curves. In this work, the maximum and minimum phase velocity for a given material will be referred to as $c_{\text{max}}$, and $c_{\text{min}}$ respectively. In addition, following from \eqref{eq:groupVelocity}, the direction of the group velocity is normal to these curves. Bécache \textit{et al.} \cite{Becache:2003ug} found that the stability of the split-field classical PML depends on the shape of slowness curves and they called this the geometrical stability condition.

\subsection{Complex stretching of the spatial coordinates }
\label{subsec:complex} 
A perfectly matched layer can be constructed by the analytic continuation of the spatial coordinate to the complex domain inside the PML region \cite{Johnson:2008wt,Chew:1994dn,Teixeira:2000vj}. Assuming that the region sufficiently far from that containing the sources and inhomogenities (see \autoref{fig:illustrationPML}) is linear and homogeneous, the radiation solution can be written as a superposition of harmonic plane waves \cite{Johnson:2008wt}. Because these waves are analytic functions of the space coordinate, the radiation solutions are also analytic and are subject to analytic continuation \cite{Teixeira:2000vj,Johnson:2008wt,Kucukcoban:2011tn}.

A coordinate transformation $ x_j \to \tilde{x}_{j}\left(x_{j}\right):\mathbb{R}\to\mathbb{C}$  is performed where $\tilde{x}_{j}\left(x_{j}\right)$ has the value of $x_{j}\thinspace$inside the physical domain and is continuous everywhere. Since homogeneity was assumed close to and inside the PML region, $x_{j}$ appears in the differential equations only as a partial derivative. Thus, the original wave equation in $x_{j}$ can be transformed into a one in $\tilde{x}_{j}$ merely by replacing $1 \mathbin{/}\partial x_{j}$ by $1 \mathbin{/}\partial \tilde{x}_{j}$. This transformed equation has the same solution in the physical domain as the original equation, but within the PML, it can be made an exponentially decaying solution with no reflections at the interface. Unfortunately, solving this differential equation along contours in the complex plane can be challenging. This can be avoided by transforming the complex coordinate back to the real coordinate $x_{j}$ \cite{Johnson:2008wt}. 

Within the PML the spatial coordinate in the PDEs only appears in the form of spatial partial derivatives. As a result, instead of defining the transformation $ x\to\tilde{x}$,  the relation between ${\partial\tilde{x}}_{j}$ and ${\partial x}_{j}$ suffices for the transformation. If the complex stretch function is defined as their ratio, i.e.,  $s_{j}\left(x_{j}\right)=\partial \tilde{x}_{j} (x_{j}) \mathbin{/} \partial x_{j}$, then 
\begin{equation}
\frac{\partial}{{\partial\tilde{x}}_{j}}=\frac{1}{s_{j}\left(x_{j}\right)}\frac{\partial}{{\partial x}_{j}}.
\label{eq:derivativeTransform}
\end{equation}
Since the stretch function is a complex function in $x_{j}$, it can be expressed in the two-parameters form: 
\begin{equation}
s_{j}\left(x_{j},\omega\right)=\alpha_{j}\left(x_{j}\right)\left[1+i\,\frac{\beta_{j}\left(x_{j}\right)}{\omega}\right],
\label{eq:CSF}
\end{equation}
where the damping coefficient, $\beta_{j}\geq0$, is responsible for damping the propagating wave inside the PML. Moreover, the scaling coefficient, $\alpha_{j}>0$, is responsible for either stretching ($\alpha_{j}>1$) or compressing ($0<\alpha<1$)the coordinate. The angular frequency, $\omega$, was added to make the damping wavevector independent. In the physical domain (see \autoref{fig:illustrationPML}) $\tilde{x}_{j}\left(x_{j}\right)=x_{j}$, so that $\beta_{j}=0$ and $\alpha_{j}=1$, whereas in the PML, $\beta_{j}>0$ and $\alpha_{j}\thinspace$can differ from 1.
\begin{figure}[h!]
\centering \includegraphics[width=1\textwidth]{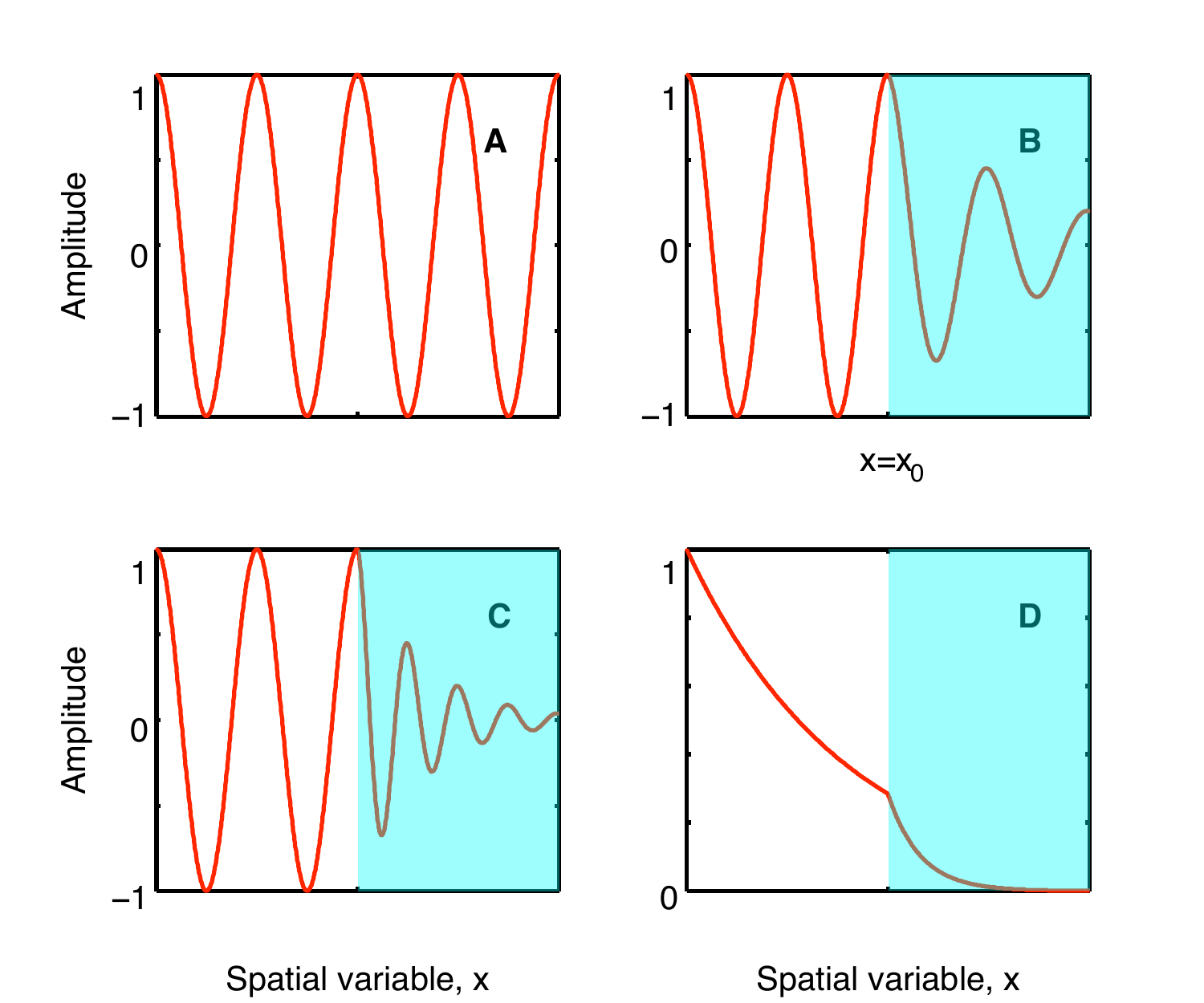} 
\caption{Illustrating the effect of complex coordinate stretching for a 1D plane wave, shown in (A), propagating into a PML. The point $x=x_{0}$ marks the beginning of the PML; the shaded region in (B), (C), and (D). (B) Shows the case where the damping coefficient $\beta\geqslant0$, and the scaling coefficient $\alpha=1$. In (C) the same value of $\beta$ as in (B) was used while $\alpha>1$. (D) Illustrating the case of an evanescent wave with $\alpha>1$.}
\label{fig:illustration1D} 
\end{figure}

To illustrate the effect of the complex coordinate stretching,  consider the simple case of the 1D oscillatory solution shown in \autoref{fig:illustration1D} (A). \autoref{fig:illustration1D}(B) shows the wave for $\beta(x)>0$ in the PML and for $\alpha(x)=1$ throughout. It can be seen  that an exponentially damped wave given is present in the PML.  \autoref{fig:illustration1D}(C) shows the cases for $\alpha(x)>1$ and the same $\beta(x)>0$ as used in (B). The real grid was stretched by $\alpha(x)$ resulting in an apparent increase in the number of cycles, which is equivalent to increasing the spatial frequency, $k$, in the original coordinate. As subsequently shown, this concept can be used to improve the discrete stability. The damping also increased in (C), since the coordinate stretching makes the wave travels more and hence, decays more. If the original wave is evanescent, the roles of $\alpha(x)$ and $\beta(x)$ are reversed. Thus, if $\alpha(x)>1$, the decaying of evanescent wave amplitude will be increased, as shown in \autoref{fig:illustration1D}(D). 

Appropriate choices are now needed for the stretch function parameters $\alpha_{j}(x_{j})$ and $\beta_{j}(x_{j})$. Despite the absence of a rigorous methodology for their choice \cite{Chew:1996wk,Kucukcoban:2011tn}, polynomial functions are often used. For the scaling coefficient, this can be expressed as 
\begin{equation}
\alpha_{j}\left(x_{j}\right)=\begin{cases}
1 & \text{if}\left|x_{j}\right|<x_{0}\\
1+\left(\tilde{\alpha}_{j}-1\right)\left(\frac{\left|x_{j}\right|-x_{0}}{d}\right)^{m} & \text{if }x_{0}\leq\left|x_{j}\right|\leq x_{0}+d,
\end{cases}\label{eq:alpha}
\end{equation}
and for the damping coefficient 
\begin{equation}
\beta_{j}\left(x_{j}\right)=\begin{cases}
0 & \text{if}\left|x_{j}\right|<x_{0}\\
\tilde{\beta}_{j}\left(\frac{\left|x_{j}\right|-x_{0}}{d}\right)^{n} & \text{if }x_{0}\leq\left|x_{j}\right|\leq x_{0}+d,
\end{cases}\label{eq:beta}
\end{equation}
where \textit{d} is the thickness of the PML, ${2x}_{0}$ is the dimension of the physical domain, which is a square centered at the origin as shown in \autoref{fig:illustrationPML}, \textit{m} and \textit{n} are the polynomial orders, and $\tilde{\alpha}_{j}$ and $\tilde{\beta}_{j}$ are constants that represent the maximum values of $\alpha$ and $\beta$ respectively. The value of $\tilde{\beta}_{j}$ can be expressed in terms of the desired amplitude reflection coefficient ($R_{j})$ due to the reflection from the outer boundary of the PML. For normal incidence, and assuming $\alpha_j=1$, it can be shown that 
\begin{equation}
\tilde{\beta}_{j}=\frac{c_{\mathrm{max}}\left(n+1\right)}{2d}\ln\left(\frac{1}{R_{j}}\right),\label{eq:reflection}
\end{equation}
where $c_{\mathrm{max}}$ is the highest wave speed which in the case of an isotropic solid, is the longitudinal wave speed. The choice of $\tilde{\alpha}_{j}$ in \eqref{eq:alpha} depends on the desired scaling (stretching or compression) of the original coordinate. The scaling of the original coordinate is simply the derivative  of the real part of $\tilde{x}_{j}$ with respect to $ x_{j}$, which is equal to $\alpha_{j}\left(x_{j}\right)$.  Hence, the value of $\tilde{\alpha}_{\mathrm{j}}$ is simply the maximum scaling of the original coordinate in the \jth direction. The orders of the polynomial functions, $m$ and $n$, in \eqref{eq:alpha} and \eqref{eq:beta} can theoretically be any integer, or even zero. Linear and quadratic polynomials are usually used, and will be used in this work unless mentioned otherwise.  

When $\alpha_{j}$ is set equal to unity in \eqref{eq:CSF}, the stretch function simplifies $s_{j}\left(x_{j},\omega\right)=1+\left[i\beta_{j}\left(x_{j}\right)\mathbin{/}\omega \right]$, which is the classical stretch function. Another form of the stretch function was introduced by Kuzuoglu and Mittra \cite{ Kuzuoglu:1996kj} who added a frequency–shift parameter $\gamma(x)$, such that $ s_j (x_j,\omega)= \alpha_j (x_j )+\left[\beta_j (x_j)\mathbin{/} (\gamma_j (x_j)-i\omega)\right]$, leading to a PML formulations that are usually called convolutional frequency shift (CFS-PML). We chose to use a two-parameter stretch function as described in \eqref{eq:CSF}. Besides terminating the evanescent waves, other advantages accrue from making $\alpha\left(x\right)\ne1$. As will be shown, it can be used to improve the stability in the PML. Moreover, the choice of $\alpha\left(x\right)>1$ can improve the absorption of near-grazing incident wave by bending the wave direction more toward the normal \cite{Komatitsch:2007bz,Zhang:2010vp,Drossaert:2007bv,Petropoulos:2000wh}. For brevity, the functional forms for $s_{j}(x_{j},\thinspace\omega),\thinspace{\thinspace\alpha}_{j}(x_{j}),\,\beta_{j}\left(x_{j}\right)$ will not be used in the remainder of this work. All other coefficients of the PDEs are assumed to be space-dependent only.

\section{Formulation of PML for elastic wave propagation}
\label{sec:formulation} 
With the help of the above background, our time-domain PML formulation can be introduced for the  wave propagation in unbounded solids. The derivation starts from the first order velocity-stress equations in the frequency domain and concludes with a second order PML time domain equations in term of the velocity field.

\subsection{Frequency domain}
\label{subsec:frequency} 
Because the stretch function $s_{j}\left(x_{j},\thinspace\omega\right)$ is a function of frequency, the PML formulation which uses complex coordinate stretching starts in the frequency domain, and then, if needed, the time domain formulation can be obtained by using the inverse Fourier transform. The frequency-domain PML equations can be obtained from Fourier transforms of \eqref{eq:waveSoild} by replacing $x$ by $\tilde{x}$, followed by the use of \eqref{eq:derivativeTransform} to transform the coordinates-stretched equations back to the original coordinates yielding: 
\begin{equation}
-\omega^{2}\thinspace\Hat{u}_{i\thinspace}s_{1}s_{2}\rho=\sum\limits _{j=1}^{2}\frac{\partial}{\partial x_{j}}\left(\sum\limits _{k,l=1}^{2}\frac{{s_{1}s_{2}C}_{ijkl\thinspace}}{{s_{j}s}_{l}}\frac{\partial\Hat{u}_{k}}{\partial x_{l}}\right).\label{eq:FreqWaveSolid}
\end{equation}
In this expression it should be noted that inside the physical domain where \mbox{$s_{1}=s_{2}=1$}, \eqref{eq:FreqWaveSolid} reduces to the frequency domain form of\eqref{eq:waveSoild}. In the PML region, \eqref{eq:FreqWaveSolid} can be looked at as the original equation but with a fictitious medium whose density is $s_{1}s_{2}\rho$ and whose elasticity tensor is $s_{1}s_{2}C_{ijkl}/s_{j}s_{l}$. Both of these coefficients are now complex and frequency dependent.

To obtain the velocity-stress formulation for the PML, we proceed in a similar manner to that used to obtain \eqref{eq:FreqWaveSolid} leading to: 
\begin{equation}
\begin{aligned}-i\omega\,\Hat{v}_{i}\,\rho s_{1}s_{2} & =\sum\limits _{j=1}^{2}{\frac{s_{1}s_{2}}{s_{j}}\frac{\partial\Hat{\sigma}_{ij}}{\partial x_{j}}}\\
-i\omega\,\Hat{\sigma}_{ij} & =\sum\limits _{k,l=1}^{2}\frac{C_{ijkl\thinspace}}{s_{l}}\frac{\partial\Hat{v}_{k}}{\partial x_{l}}\,.
\end{aligned}
\label{eq:FreqStressVelocity}
\end{equation}
Either \eqref{eq:FreqWaveSolid} or \eqref{eq:FreqStressVelocity} can be be used for frequency domain simulation. In addition, \eqref{eq:FreqStressVelocity} will be used in the next section to obtain the time domain equations. 

\subsection{ Time-domain formulation}
\label{subsec:timeDomain} 
We proceed by first splitting each of the stress field components in \eqref{eq:FreqStressVelocity} into two non-physical components, $\sigma_{ij}^{1}$ and $\sigma_{ij}^{2}$, while keeping the velocity field components unsplit. Since ${s_{1}s_{2}}/s_{j}$ in \eqref{eq:FreqStressVelocity} does not depend on $x_{j}$ it can be placed inside the $x_{j}$ derivative, leading to 
\begin{equation}
\begin{aligned}-i\omega\Hat{v}_{i}\rho\thinspace s_{1}s_{2} & =\sum\limits _{j=1}^{2\thinspace}{\thinspace\thinspace\frac{\partial}{\partial x_{j}}\thinspace}\left(\frac{s_{1}s_{2}}{s_{j}}\sum\limits _{l=1}^{2}\Hat{\sigma}_{ij}^{l}\right)\\
-i\omega\Hat{\sigma}_{ij}^{l} & =\sum\limits _{k=1}^{2}\frac{C_{ijkl\thinspace}}{s_{l}\thinspace}\thinspace\frac{\partial\Hat{v}_{k}}{\partial x_{l}}.
\end{aligned}
\label{eq:derevationElastic1}
\end{equation}
Multiplying the first by $-i\omega$, the second by ${s}_{l}$, and expanding $s_{1}$ and $s_{2}$ using \eqref{eq:CSF}, results in
\begin{equation}
\begin{aligned}\rho\left[\left(-i\omega\right)^{2}+(-i\omega)\left(\beta_{2}+\beta_{1}\right)+\beta_{1}\beta_{2}\right]\Hat{v}_{i} & =\sum\limits _{j=1}^{2\thinspace}{\frac{1}{\alpha_{j}}\thinspace\frac{\partial}{{\partial x}_{j}}\left[\left(-i\omega+\frac{\beta_{1}\beta_{2}}{\beta_{j}}\right)\sum\limits _{l=1}^{2}{\thinspace\Hat{\sigma}_{ij}^{l}}\right]}\\
\left(-i\omega\right)\thinspace\Hat{\sigma}_{ij}^{l}+\beta_{l}\Hat{\sigma}_{ij}^{l} & =\sum\limits _{k=1}^{2}\frac{C_{ijkl\thinspace}}{\alpha_{l}}\frac{\partial\Hat{v}_{k}}{\partial x_{l}}.
\end{aligned}
\label{eq:derevationElastic2}
\end{equation}

The time domain form of \eqref{eq:derevationElastic2} can now be obtained by taking its inverse Fourier transform without a need for convolution $(-i\omega\Rightarrow\thinspace\partial/\partial t)$, leading to 
\begin{equation}
\begin{aligned}\rho\left[\frac{\partial^{2}v_{i}}{\partial t^{2}}+\left(\beta_{2}+\beta_{1}\right)\frac{\partial v_{i}}{\partial t}+\beta_{1}\beta_{2}v_{i}\right] & =\sum\limits _{j=1}^{2}{\frac{1}{\alpha_{j}}\frac{\partial}{\partial x_{j}}\left[\sum\limits _{l=1}^{2}\left(\frac{\partial\sigma_{ij}^{l}}{\partial t}+\frac{\beta_{1}\beta_{2}}{\beta_{j}}\sigma_{ij}^{l}\right)\right]}\\
\frac{\partial\sigma_{ij}^{l}}{\partial t}+\beta_{l}\sigma_{ij}^{l} & =\sum_{k=1}^{2}\frac{C_{ijkl}}{\alpha_{l}}\frac{\partial v_{k}}{\partial x_{l}}.
\end{aligned}
\label{eq:derevationElastic3}
\end{equation}
By substituting $\partial\sigma_{ij}^{l}/\partial t$ from the second to the first and simplifying, yields 
\begin{equation}
\begin{aligned}\rho\left[\frac{\partial^{2}v_{i}}{\partial t^{2}}+\left(\beta_{1}+\beta_{2}\right)\frac{\partial v_{i}}{\partial t}+\beta_{1}\beta_{2}v_{i}\right] & =\sum\limits _{j=1}^{2}\frac{1}{\alpha_{j}}\frac{\partial}{{\partial x}_{j}}\left[\sum\limits _{k,l=1}^{2}\frac{C_{ijkl}}{\alpha_{l}}\frac{\partial v_{k}}{\partial x_{l}}+\sum\limits _{l=1}^{2}\left(\frac{\beta_{1}\beta_{2}}{\beta_{j}}-\beta_{l}\right)\sigma_{ij}^{l}\right]\\
\frac{\partial\sigma_{ij}^{l}}{\partial t}+\beta_{l}\sigma_{ij}^{l} & =\sum_{k=1}^{2}\frac{C_{ijkl}}{\alpha_{l}}\frac{\partial v_{k}}{\partial x_{l}}.
\end{aligned}
\label{eq:derevationElastic4}
\end{equation}
Noting that if $j\ne l$, then $\left(\beta_{1}\beta_{2} \mathbin{/}\beta_{j}\right)-\beta_{l}=\beta_{l}-\beta_{l}=0$ so that only four of the eight split stress components $\left(\sigma_{ij}^l\right)$, namely $\sigma_{ij}^{j}$ remain in the first equation. These four non-physical split stress components are needed to solve for the velocity field and will be considered as auxiliary variables denoted by $\sigma_{ij}^{j}\equiv A_{ij}$. Thus, our time domain PML formulation consists of two second-order velocity field equations and four auxiliary equations that can be expressed as 
\begin{equation}
\begin{aligned} 
\tilde{\rho}\left(\frac{\partial^{2}v_{i}}{\partial t^{2}}+b\,\frac{\partial v_{i}}{\partial t}+c \,v_{i}\right) & =\sum\limits _{j=1}^{2}\frac{\partial}{{\partial x}_{j}}\left[ \left(\sum\limits _{k,l=1}^{2}\tilde{C}_{ijkl}\frac{\partial v_{k}}{\partial x_{l}}\right)+ a_j A_{ij}\right]\\
\frac{\partial A_{ij}}{\partial t}+\beta_{j}A_{ij} & =\sum_{k=1}^{2}\frac{C_{ijkj}}{\alpha_{j}}\frac{\partial v_{k}}{\partial x_{j}},
\end{aligned}
\label{eq:pmlElastic}
\end{equation}
where $\tilde{\rho}=\alpha_1 \alpha_2 \rho$, $\tilde{C}_{ijkl}= \alpha_1 \alpha_2 C_{ijkl}\mathbin{/}\alpha_j \alpha_l$, $a_j= \left(\alpha_{1}\alpha_{2}\mathbin{/}\alpha_{j}\right)\left[ \left(\beta_{1}\beta_{2}\mathbin{/}\beta_{j}\right) -\beta_{j}\right]$, $b=\beta_1+\beta_2$, and $c=\beta_1 \beta_2$. It should be noted that the number of equations in \eqref{eq:pmlElastic} is less than that present in the classical form and the convolutional form (typically 10 and 13 equations respectively \cite{Komatitsch:2007bz}). Other time domain PML formulations follow a similar pattern.  

If preferred, a set of displacement time domain PML equations can readily be obtained by integrating \eqref{eq:pmlElastic} with respect to time. Since the coefficients of \eqref{eq:pmlElastic} are time independent and $A_{ij}$ is only an auxiliary variable, this results in equations of the same form as the above equations but with the velocity field, $v_{i}$ replaced by the displacement field, $u_{i}$. It should be noted that, in the physical domain, the two equations in \eqref{eq:pmlElastic} are decoupled, and the displacement form of the first one is identical to the original equation, \eqref{eq:waveSoild}, which should be the case for any valid PML formulation.

\section{Stability in the PML}
\label{subsec:stability} 
In general, when $m,\thinspace n\ne0$ in \eqref{eq:alpha} and\eqref{eq:beta}, \eqref{eq:pmlElastic} is a variable coefficient PDE in the PML. However, to study the stability of the variable coefficient problem, it is helpful to assume constant coefficients, which allows use of the plane wave analysis approach \cite{Duru:2012va,Appelo:2006wz,Becache:2003ug}. In the physical domain, we know that the roots of the characteristic polynomial are real and there is no stability issue, but in the PML complex roots can be present leading to potential instability. When $\omega$ is complex, the plane wave solution, as given by \eqref{eq:harmonicWave}, becomes $\mathbf{u}=\mathbf{u}_{0}\, e^{\Im\left\{ \omega\right\} t}\, e^{i\left(\mathbf{k\cdot x}-\Re\left\{ \omega\right\} t\right)}$. Thus, the sign of the imaginary part of $\omega$ determines the stability of \eqref{eq:pmlElastic}. Specifically, if $\Im\left\{ \omega\right\} >0$, the solution grows exponentially with time, alternatively if 
\begin{equation}
\Im\left\{ \omega\left(\mathbf{k}\right)\right\} \leqslant0,\quad\forall\,\mathbf{k}\in\mathbb{R}^{2},\label{eq:imaginaryOmega}
\end{equation}
then \eqref{eq:pmlElastic} is stable.

Numerical results and studies \cite{Appelo:2006wz,Becache:2003ug,Daros:2007de} have shown that instability starts in in one or both directions of the PML, but not in the corner region where the full PML equation is involved. Just one direction for the stability analysis will be considered, namely, the $x_{1}$ direction, where $\beta_{2}=0$ and $\alpha_{2}=1$. For this case the 8$^{\mathrm{th}}$ order characteristic polynomial of \eqref{eq:pmlElastic} is
\begin{equation}
F{}_{1}\left(\omega,\thinspace k_{1},k_{2},\beta_{1},\alpha_{1}\right)\equiv F_{0}\left[\left(\omega+i\beta_{1}\right)\omega,\frac{k_{1}}{\alpha_{1}}\thinspace\omega,\thinspace k_{2}\left(\omega+i\beta_{1}\right)\right]=0.
\label{eq:dispersionPML}
\end{equation}
where $F_{0}$ is defined by \eqref{eq:ellipticityConditions}. Assuming $\omega=i\eta$, which makes \eqref{eq:dispersionPML} a real-coefficient 8$^{\mathrm{th}}$ order polynomial in $\eta$ and, according to the complex conjugate root theorem its roots, $\eta\left(k_{1},k_{2},\beta_{1},\alpha_{1}\right)$, come in complex conjugate pairs. Hence
\begin{lm}, \label{th:pair}
The roots of \eqref{eq:dispersionPML}, $\omega\left(k_{1},k_{2},\beta_{1},\alpha_{1}\right)$,
come in pairs:  each pair has the same imaginary part and the real parts differ only in sign.
\end{lm}
If none of the four pair of roots of \eqref{eq:dispersionPML} has a positive imaginary part, stability in the $x_{1}$ direction of the PML is assured. 

First, consider the case in which $\alpha_{1}=1\thinspace$. For this case \eqref{eq:dispersionPML} is identical to the equation for $\tilde{F}_{pml}$ as given by Bécache \textit{et al} \cite{Becache:2003ug} as part of the dispersion relation of the classical split-field PML (see their equation (64)). Using the perturbation techniques, they studied the stability of $F_{1}\left(\omega,\thinspace k_{1},k_{2},\beta_{1},1\right)$ and found, among other results,  the following:
\begin{enumerate}
\item All the necessary and sufficient stability conditions could be expressed in terms of the elasticity coefficients.

\item High frequency  \emph{stability geometric} condition  (Theorem 2 of their work):\\
It is necessary that  all points on the slowness curve  satisfy
\begin{equation}
S_{j}{\times\left(V_{g}\right)}_{j}\geqslant0,\label{eq:geometricStability}
\end{equation}
for the PML in the $x_{j}$ direction to be stable. This means that the \jth component of the group velocity is in the same direction as the \jth component of the slowness vector, which can be readily identified on the slowness curves shown in \autoref{fig:slowness}. Violating this condition usually causes the most severe instability. The geometric stability was also found to be necessary condition for other PML formulations \cite{Appelo:2006wz,Duru:2012va}.
\item Because of the symmetries in the orthotropic media, it is enough to consider the first quarter of the $\mathbf{k}-$space ($k_1 \mathbin{>} 0$ and  $k_2 \mathbin{>} 0$). 
\end{enumerate}

To find the stability condition for our PML formulation, we need to consider \eqref{eq:dispersionPML} with the general case of $\alpha_{1}\neq1$. By inspection, it is evident that the roots of $F_{1}\left(\omega,\alpha_{1}k_{1}, k_{2},\beta_{1},\alpha_{1}\right)=0$ are the same as the roots of $F_{1}\left(\omega,k_{1},k_{2},\beta_{1} 1\right)$=0, hence,
\begin{co}
\label{th:same}
Changing the scaling parameter, $\alpha_1$ from unity will cause any root of $F_1=0$ to be moved in $k-$space. For the continuous case, such a movement can never cause any unstable roots to become stable. Therefore, the necessary and sufficient condition for the stability of our constant coefficient continuous problem, as defined  by \eqref{eq:pmlElastic}, are exactly the same as the ones reported by Bécache \textit{et al} \cite{Becache:2003ug} for their split-field system.
\end{co}

\begin{figure}[h!]
\centering \includegraphics[width=1\textwidth]{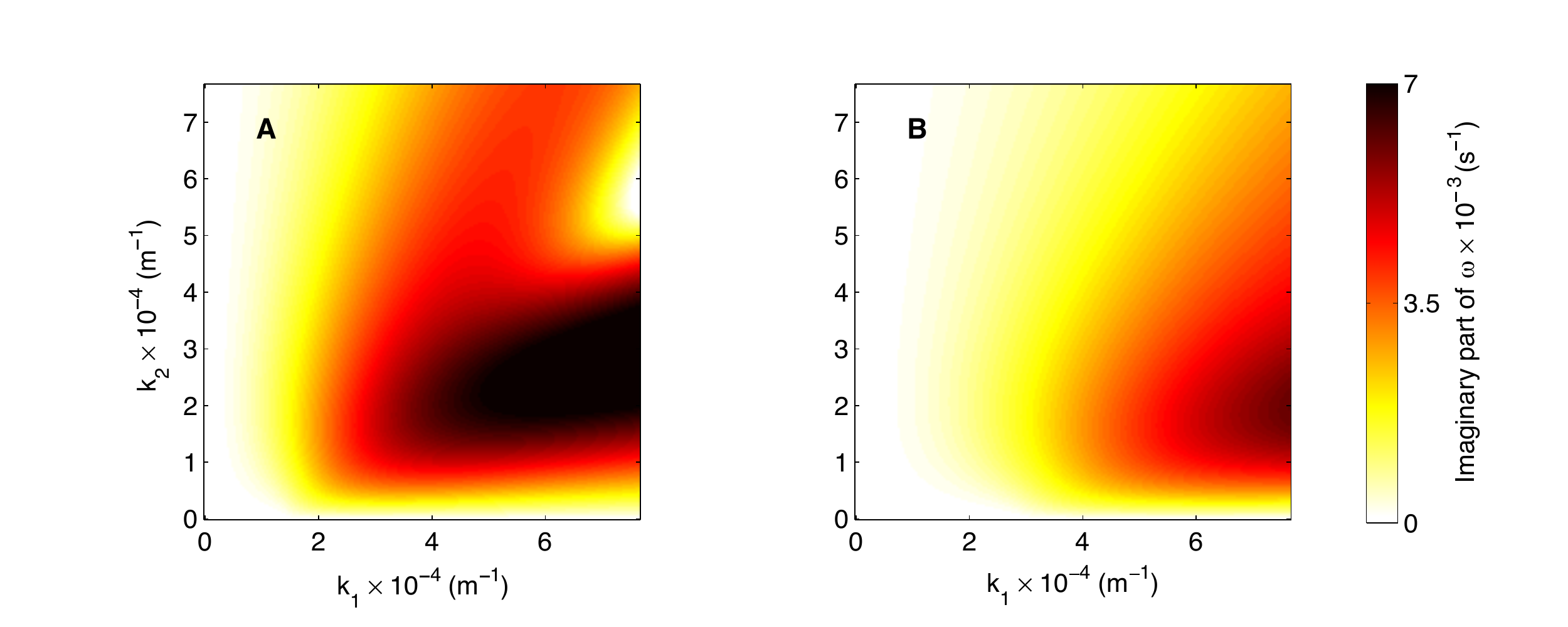}
\caption{Illustrating the effect of incorporation of the scaling coefficient $\alpha_{1}$ on the roots of \eqref{eq:dispersionPML}, for the  continuous, constant coefficient problem. The color maps show the imaginary part of the unstable pair of roots, $\Im\left\{\omega\right\} $, for material III. (A) For the classical case of $\alpha_{1}=1$. (B) For a case in which $\alpha_{1}=2$. The roots merely shifted to higher wavevectors.}
\label{fig:imaginaryOmega} 
\end{figure}

\autoref{fig:imaginaryOmega} shows the effect of increasing the scaling parameter on the roots of \eqref{eq:dispersionPML} for material III in \autoref{tab:material}. In (A), the imaginary part of the unstable pair of roots are shown for a range of wavevectors in the first quarter of the $\mathbf{k}-$space,  for the case of $\alpha_1 =1$. In (B), the same pair of roots is plotted over the same range of wavevectors but for the case of $\alpha_1 =2$. Indeed, as suggested in Corollary \autoref{th:same}, the roots were just shifted.  

Corollary \autoref{th:same} shows that incorporating the scaling parameter will not improve our continuous constant coefficient problem in \eqref{eq:pmlElastic}. Though, since PML is meant to be used for numerical simulation, the  more relevant question is whether the  stability of the discrete problem  that corresponds to an unstable continuous problem can be improved?  In fact, this was shown to be case  if the unstable continuous modes  are not well resoled by the discrete mesh \cite{Duru:2012va, Appelo:2006wz, Kreiss:2013tb}, specially for second order formulations \cite{Kreiss:2013tb}.
 
If the unstable modes of the PML formulation shown by \eqref{eq:pmlElastic} were in higher wavevector range than can be resolved by the mesh, then our discrete model could be  expected to stable. On the other hand, if the unstable continuous modes are resolvable, increasing $\alpha_1$ shifts the modes to higher wavevectors which might improve the discrete stability. This will be the case if the modes of the lower wavevector, which now cover the resolvable range, have a smaller imaginary part. To investigate this, we return to  the dispersion relation given by \eqref{eq:dispersionPML} and let $\xi=k_1\mathbin{/} \alpha_1$ (remember $\alpha_1 >1$). For a fixed value of $\beta_1$, the roots of the dispersion relation are continuous functions in term of  $\xi$, and $k_2$ thanks to  the            \emph{implicit function theorem}. Noting that decreasing the value of $\xi$ is equivalent to  increasing $\alpha_1$ or decreasing $k_1$, as $ \xi \to 0 $ the dispersion relations becomes:
\begin{equation}\label{eq:k=0}
\left(\omega+i\beta_1\right)^4 \left(\omega^2-C_{22}k_2^2\right) \left(\omega^2-C_{33}k_2^2\right) =0,
\end{equation} 
which admits no solution, $\omega(k_2)$, with a positive imaginary part. in fact, two  of the four pairs of roots of \eqref{eq:k=0} have the imaginary parts equal to $- \beta_1$, while the imaginary parts  of the other two pairs are equal to zero. Since the root of the dispersion relation are continuous functions in term of $\xi=k_1/\alpha_1$, it follows that:
\begin{tr}
\label{th:alpha}
By increasing $\alpha_1$ beyond a certain threshold, the discrete stability of \eqref{eq:pmlElastic} starts to improve. 
\end{tr} 
In fact, the results given in \autoref{subsec:anisotropic} provide evidence that supports Theorem~\autoref{th:alpha}

\section{Numerical Methods and Results }
\label{sec:numerical} 
In all our discrete studies, the source of excitation was a 1~mm diameter infinite cylinder embedded in an infinite 2D medium. To model the infinite medium we assumed a physical domain of 1.0 cm$^{2}$ surrounded by a 1.0~mm PML. The boundary of the cylinder was assumed to vibrate normally (unless mentioned otherwise) with a velocity, whose normalized time-dependence is given by the first derivative of a Gaussian, i.e.,
\begin{equation}
v_{0}\left(t\right)=-\sqrt{2e}\thinspace\pi f_{0}\left(t-t_{0}\right)\thinspace e^{-\pi^{2}f_{0}^{2}\left(t-t_{0}\right)}\label{eq:source}
\end{equation}
where $f_{0}$ is the dominant frequency and $t_{0}$ is a source delay time. For all numerical experiments $f_{0}=1500\thinspace$Hz and $t_{0}=1$~ms. 90{\%} of the energy of the signal is contained below the frequency $f_{c}=1900$ Hz.

COMSOL Multiphysics was used in combination with MATLAB to numerically solve \eqref{eq:pmlElastic} using the finite element method. Dirichlet boundary conditions were used throughout: specifically, $\mathrm{\mathbf{v}}=v_{0}\left(t\right)\mathbf{\Hat{n}}$ on the surface of the cylinder and $\mathrm{\mathbf{v}}=0$ on the outer boundary of the computational domain, where $\mathbf{\Hat{n}}$ is the normal unit vector to cylinder surface. A square mesh was used for the PML region, but we retained a triangular shape in the physical domain. The choice of an appropriate mesh size is governed by the shortest wavelength of significance for the propagating pulse, i.e., $c_{\text{min}}/f_{c}$. Since a second order shape function was used in our finite element method the mesh size was taken to be $h_{0}=\frac{1}{5}\left(c_{\text{min}}/f_{c}\right)$, which corresponds to ten degrees of freedom per wavelength. For time discretization we used an implicit method, specifically the generalized alpha method. Compared to explicit methods the stability of implicit methods is not as sensitive to the choice of the time step, time step size of just less than $h_{0}/c_{\text{max}}$ was used, which is sufficient to make optimal use of the mesh.

\subsection{Model validation}
\label{subsec:model}
\begin{figure}[h!]
\centering \includegraphics[width=1\textwidth]{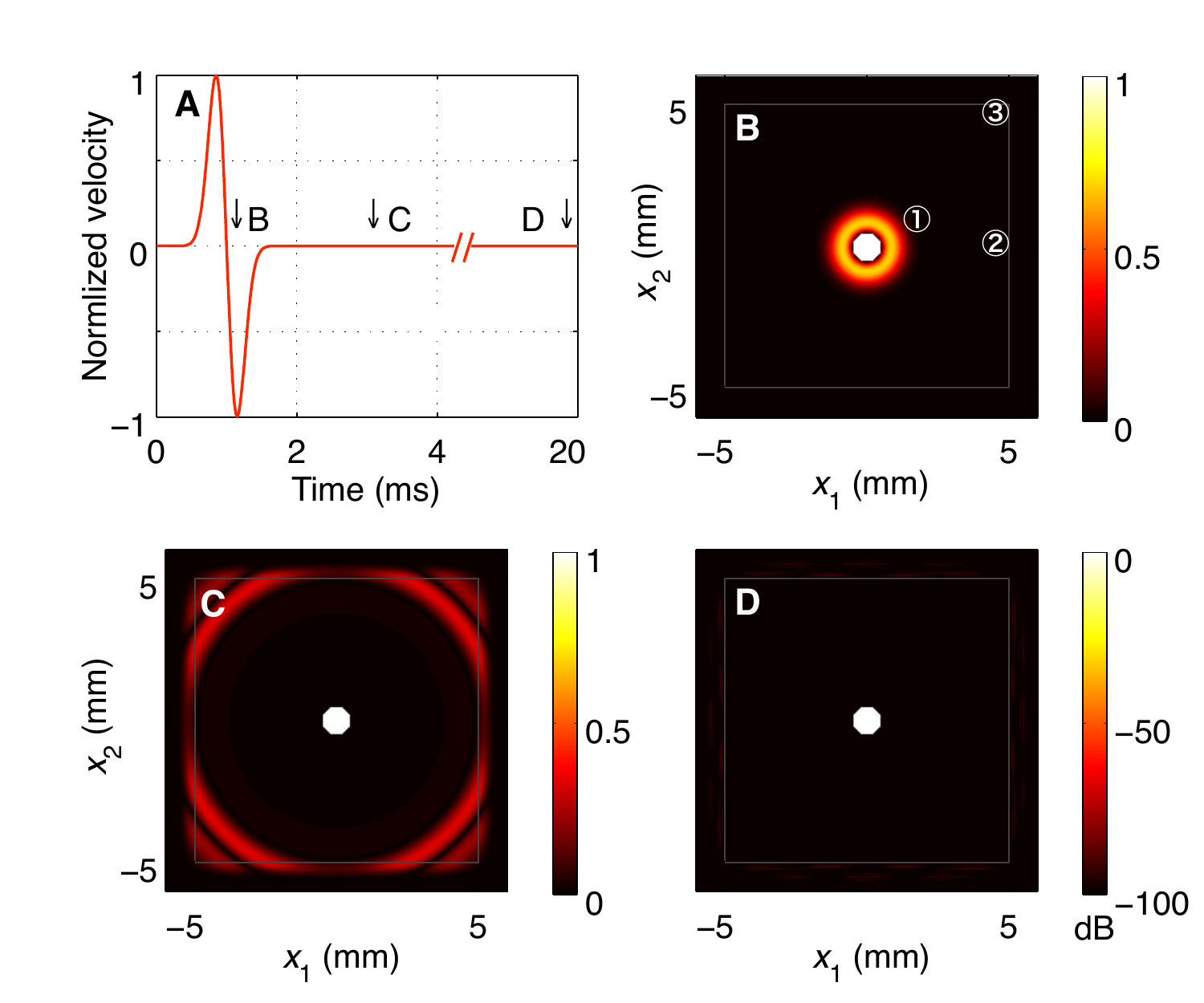}
\caption{(B), (C) and (D) are snapshot images showing the amplitude of the particle velocity for a transient longitudinal wave propagating in the isotropic solid medium listed in \autoref{tab:material}. The radiation originates from a surface of a 1 mm diameter cylinder that radial with the velocity profile shown in (A). Marked on the time axis of (A) are the times at which the snapshots in (B), (C) and (D) are taken. Note that (B) and (C) have linear scales, while (D) is in dB's. The points \ding{192}, \ding{193}, and \ding{194} in (B) are in the physical domain where the solutions are compared to the analytical solutions in \autoref{fig:validation}.}
\label{fig:isotropicFEM} 
\end{figure}

\begin{figure}[h!]
\centering \includegraphics[width=1\textwidth]{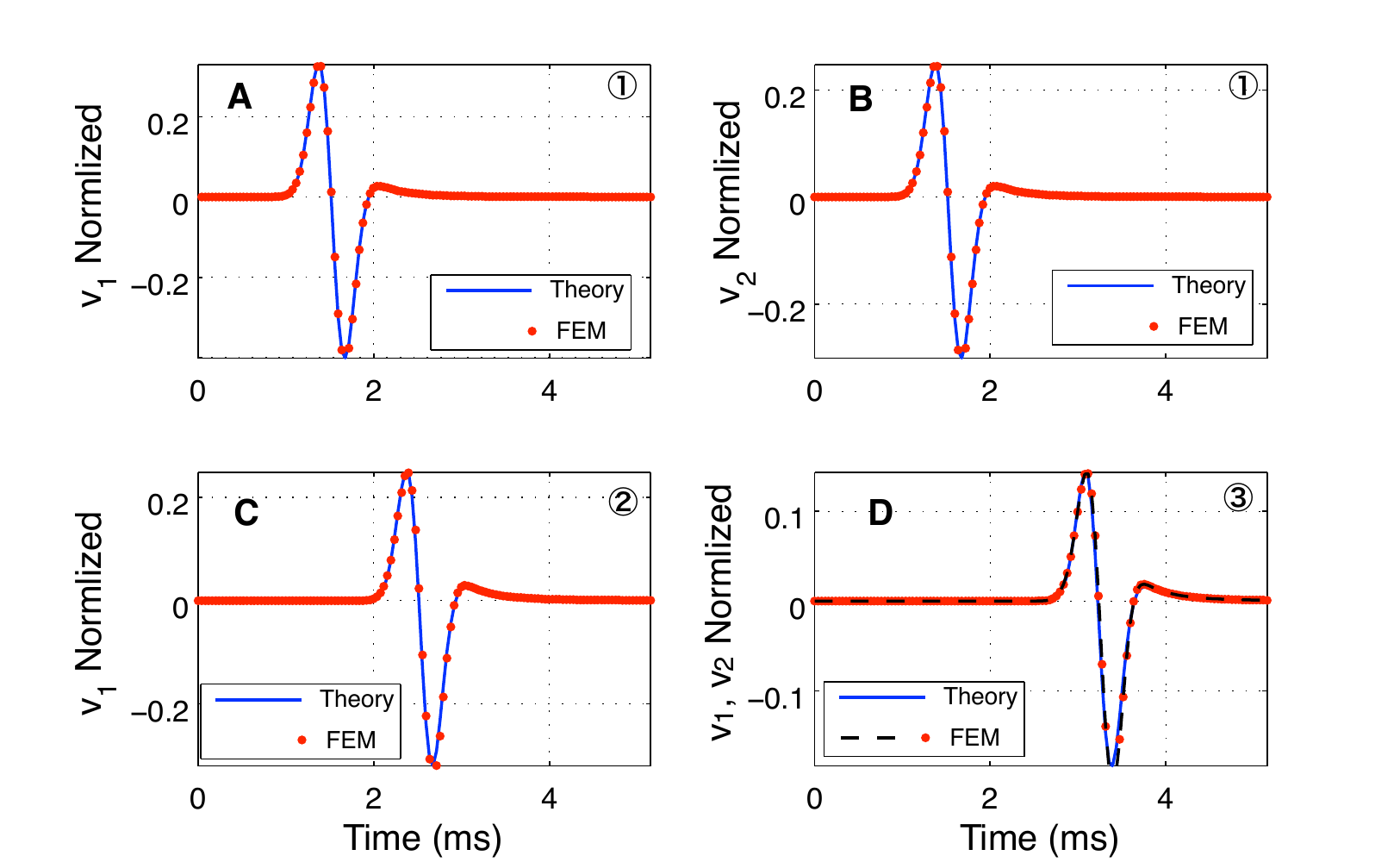} 
\caption{Validation results: the three points \ding{192}, \ding{193}, and \ding{194} marked on \autoref{fig:isotropicFEM} (B) are the locations in the physical domain where the particle velocities were both simulated and analytically calculated. The solid line is the theoretical and the dashed line is from the FEM simulation. (A) and (B) show the two components of the velocity field at point \ding{192}. (C) Velocity field at point \ding{193}. (D) Showing both components of the velocity field at point \ding{194}.}
\label{fig:validation} 
\end{figure}

Simulation of wave propagation in unbounded isotropic solid is presented in \autoref{fig:isotropicFEM} where snapshots of the propagation pulse described by \eqref{eq:source} are shown for three instants of time. To test the accuracy with which these simulations describe the propagating pulse, we made use of the exact solution for a monochromatic compressional wave caused by an infinitely long vibrating cylinder in an unbounded isotropic solid \cite{Beltzer:1988ut}. By multiplying this with the Fourier transform of \eqref{eq:source}, then taking the inverse Fourier transform the time-domain analytical solution was obtained and compared to the FEM results. As shown in \autoref{fig:validation} the agreement is excellent, thereby providing good evidence for the effectiveness of our PML formulation in simulating unbounded media and the correctness of the FEM model.

\begin{figure}[t]
\centering \includegraphics[width=1\textwidth]{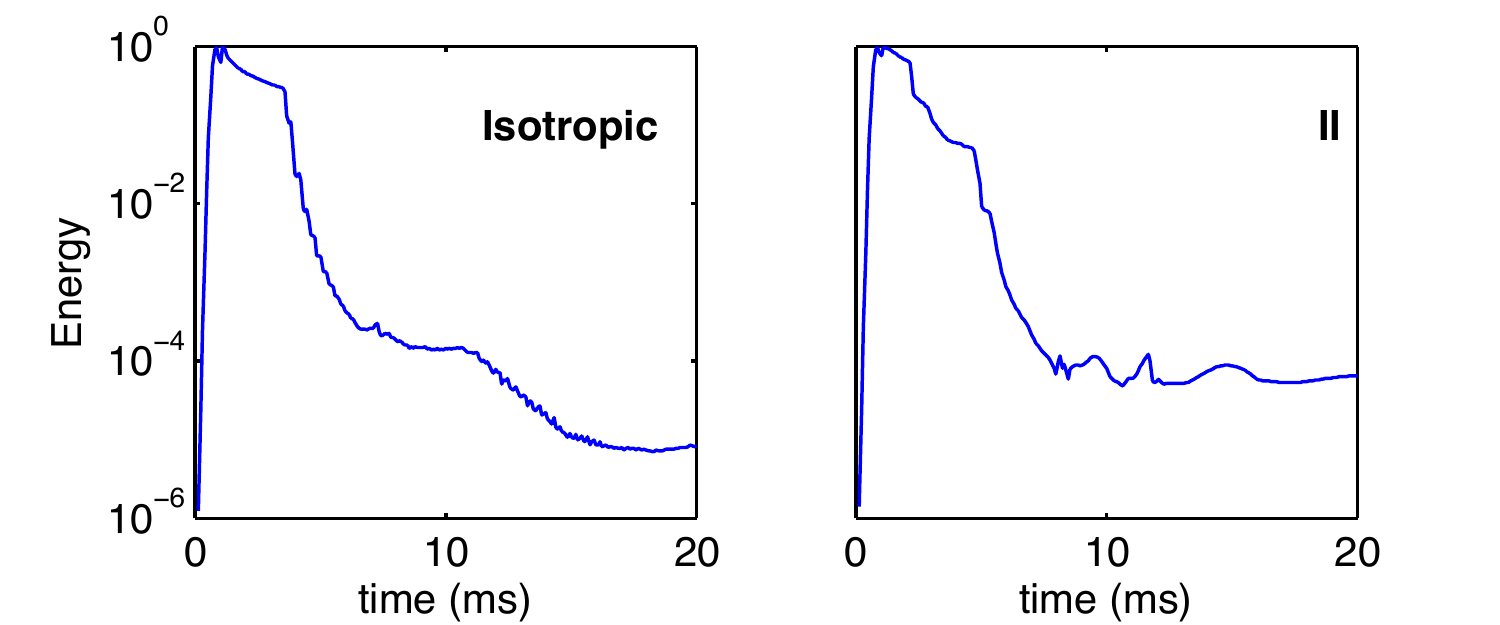} 
\caption{Showing the evolution of energy in the physical domain, as represented by $\|\sqrt{v_{1}^{2}+v_{2}^{2}}\|_{\infty}$ for the isotropic material and material II.}
\label{fig:energyIsotropic} 
\end{figure}

Another measure of the effectiveness of the PML can be obtained by looking at the manner in which the energy in the physical domain evolves in time to ensure that no energy is reflected back into the physical domain. There are several ways of doing this \cite{Duru:2012va,Li:2010tu,Appelo:2006wz}, one of which is to calculate the maximum magnitude of the particle velocity in the physical domain $\left\Vert \sqrt{v_{1}^{2}+v_{2}^{2}}\right\Vert _{\infty}$, and to see how this evolves in time. This is shown in \autoref{fig:energyIsotropic} for the isotropic material as well as for material II, both of which have no stability issues. The discrepancies in the energy curve is due to the fact that $\left\Vert \sqrt{v_{1}^{2}+v_{2}^{2}}\right\Vert _{\infty}$ is a local measure at the maximum-valued point, and not an averaged measure over the whole physical domain like other norms, which on the other hand makes it more sensitive measure to any reflection.

\subsection{Anisotropic solid: stability }
\label{subsec:anisotropic} 
The last  three materials in \autoref{tab:material} violate the stability conditions  as described by  by Bécache \textit{et al} \cite{Becache:2003ug}. For these, the plane wave analysis was used in order to study the stability. This approach assumed that all the coefficients of the PDE, including $\alpha_{j}$ and $\beta_{j}$ are constant throughout the PML. In spite of these assumptions, the plane wave analysis provides a valuable guide for achieving stability in the discrete variable-coefficients problem \cite{Duru:2012va,Appelo:2006wz,Becache:2003ug}.

\subsubsection{Plane wave analysis results}
\label{subsubsec:plane}
The imaginary parts of the roots, $\Im\left\{ \omega\left(\mathbf{k}\right)\right\} $, of \eqref{eq:dispersionPML} were numerically obtained, using MATLAB, over a range of wavevectors appropriate to our analysis. Since the materials being considered are orthotropic, it is sufficient to study the first quarter of the $\mathbf{k}-$space  \cite{Becache:2003ug}. As discussed earlier, the stretch function parameters were assumed to be constants. For all cases, $\beta_{1}$ that corresponds to a reflection coefficient  $R_{1}=1\times{10}^{-6}$ was used. 
 
\begin{figure}[!h]
\centering \includegraphics[width=1\textwidth]{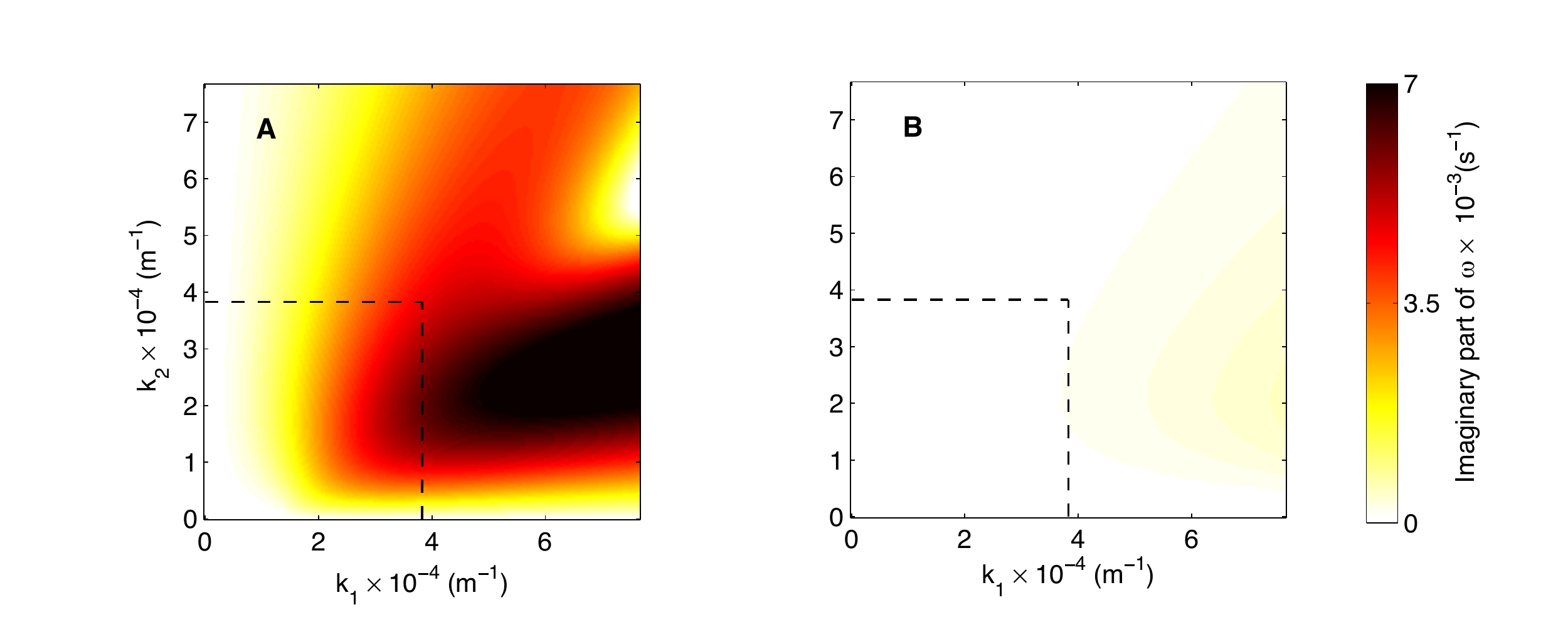} 
\caption{Illustrating the effect of incorporation of the scaling coefficient
$\alpha_{1}$ on the discrete stability. The color maps show the continuous imaginary part
of the unstable pair of roots, $\Im\left\{ \omega\right\} $, for material III. (A) For  $\alpha_{1}=1$. (B) For $\alpha_{1}=10$. The dashed lines indicate the highest wavevectors that can be resolved for the mesh size used in the discrete simulations (see text for details).}
\label{fig:imaginaryOmega2} 
\end{figure}

Material III is the most challenging in terms of stability \cite{Duru:2012va,Li:2010tu,Appelo:2006wz,Becache:2003ug} since it severely violates the geometric stability as expressed in \eqref{eq:geometricStability}. This is evident from the slowness curve of \autoref{fig:slowness}. For this material the effect of coordinate stretching, making $\alpha_{1}>1$, was  examined in detail and reported in \autoref{fig:imaginaryOmega} which was discussed in \autoref{subsec:stability}, and \autoref{fig:imaginaryOmega2} which will be discussed below.

\autoref{fig:imaginaryOmega2} contain two panels each of which shows the imaginary part of the unstable pair of roots of \eqref{eq:dispersionPML}. Panel (A) corresponds to using the classical stretch function, $\alpha_{1}=1$, while in (B) $\alpha_{1}=10$ was used. As one would expect, in (B) the roots were shifted to even higher wavevectors than in the case of $\alpha_{1}=2$ in \autoref{fig:imaginaryOmega} (B). Though, the continuous problem still unstable because the positive imaginary part only shifted. But our interest is in discrete solutions so that the question now arises as to what would be the effect of this shift
on the discrete problem.

To answer this question, we note that the highest spatial frequency that can be numerically resolved in each direction is $\pi\mathbin{/}h_{0}$. Dashed lines are included in both graphs of \autoref{fig:imaginaryOmega2} to represent this threshold. It is clear from (A) that unstable roots with positive imaginary part are present in the wavevectors range that can be resolved by discrete models, i.e., below the dashed lines. Hence, we expect the FEM simulations to be unstable for this case. On the other hand in (B), the unstable roots are shifted beyond the wavevectors range that can be discretely resolved. Therefore substantial increase in the stability of the FEM simulations is expected. Similar results were also obtained for the $x_{2}$ direction but, because the violation in the $x_{2}$ direction for this material is very severe a higher value for $\alpha_{2}$ was needed to ensure stability over the same range of wavevectors.

Similar plane-wave analyses were performed for materials IV, and V. For material IV, even with $\alpha_{1}=1$, the unstable pair of roots were found to occur at higher wavevectors than those that can be numerically resolved and hence, these should be stable in the FEM simulations. For material V, the unstable pair were below the dashed line over for $\alpha_{1}=1$, suggesting the possibility of a numerical instability.

\subsubsection{Finite element results}
\label{subsubsec:finite} 
\begin{figure}[h!]
\centering \includegraphics[width=1\textwidth]{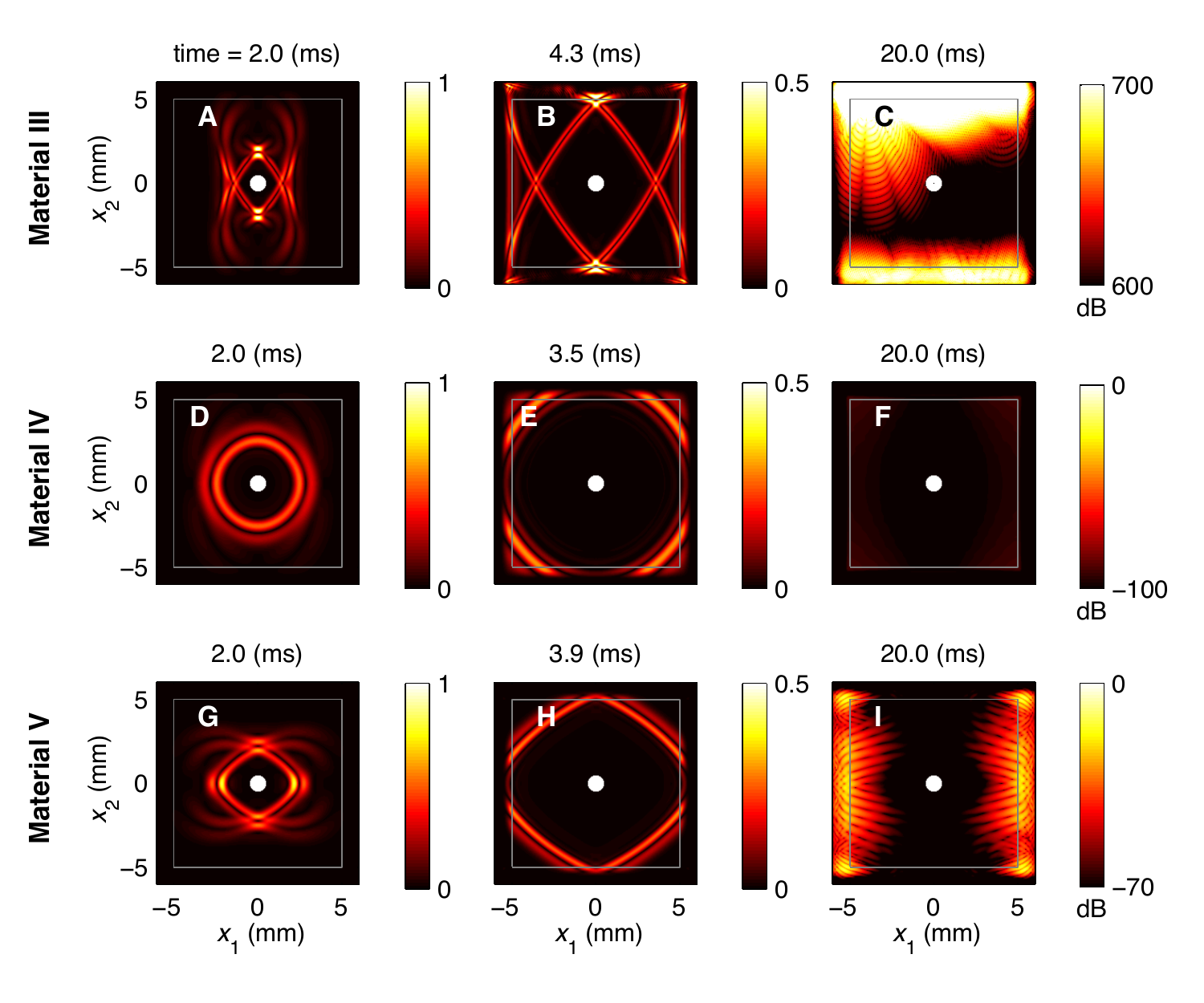} 
\caption{Snapshot images showing the waveforms, originating from same cylinder as shown in \autoref{fig:isotropicFEM}, but propagating in three different anisotropic solid media, namely III, IV, and V as specified in \autoref{tab:material}. The middle column snapshot times were chosen to approximately correspond to the quasi-shear wave being absorbed by the PML. The color maps on the third column are in decibel scale.}
\label{fig:anisotropicFEM1} 
\end{figure}

For the discrete FEM simulation, $\alpha_{j}$ and $\beta_{j}$ are not constants, rather they are functions of $x_{j}$ as shown in \eqref{eq:alpha} and \eqref{eq:beta}. Since the unstable modes are usually the quasi-shear modes \cite{Becache:2003ug}, the media was excited by tangential vibrations of the cylinder surface in order to have most of the wave energy in that mode. \autoref{fig:anisotropicFEM1} shows the FEM result for the three unstable materials using the classical stretch function, i.e., without introducing any scaling coefficients. This was achieved by setting $\tilde{\alpha}_{j}=1$ in \eqref{eq:alpha}. In \eqref{eq:reflection} the reflection coefficients were chosen to be $R_{j}=1.{10}^{-6}$, and in \eqref{eq:alpha} and \eqref{eq:beta} $m=n=2$ were used. Each row in this figure shows three snapshots for the wave propagating in materials III, IV, and V, respectively. In the last column, to better show the amount of energy that remains in the computational domain, a dB scale has been used. As expected from the plane wave analysis \autoref{fig:anisotropicFEM1} (F) shows that even after a long time ( 20 ms), material IV is stable. On the other hand, for material V, as shown in (I), some instabilities have emerged in PML region. Material III shows serious instabilities that appear to start after the arrival of the slow wave to the PML region ($\sim 4$ ms).

\begin{figure}[h!]
\centering \includegraphics[width=1\textwidth]{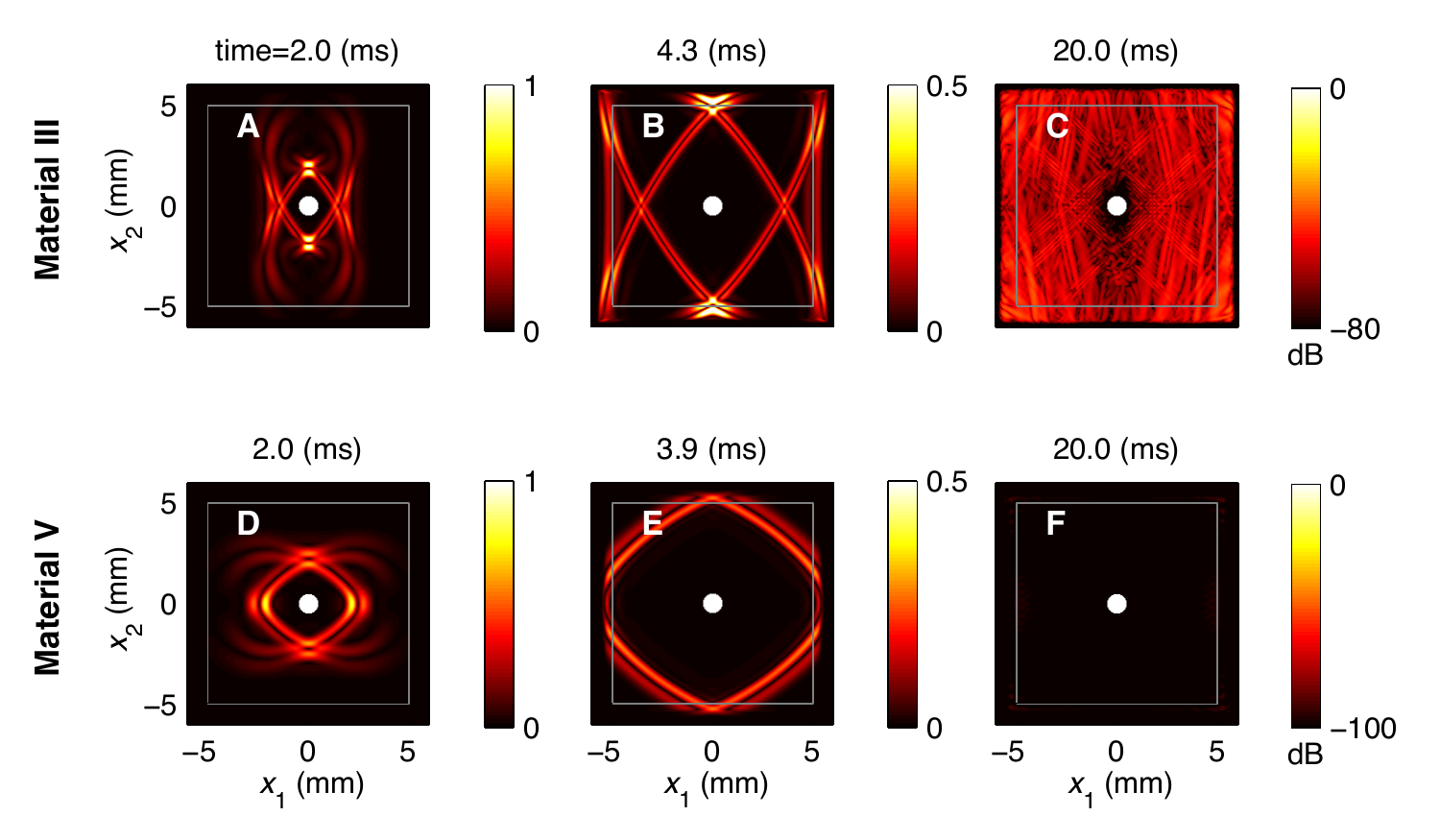} 
\caption{Propagation snapshots as in \autoref{fig:anisotropicFEM1}, but just for materials III and V, after introducing the scaling coefficient, $\tilde{\alpha}_{j}$. For material III $\tilde{\alpha}_{1}=20$ and $\tilde{\alpha}_{2}=90$. For material V $\tilde{\alpha}_{1}=10$ and $\tilde{\alpha}_{2}=1$. Comparison with \autoref{fig:anisotropicFEM1} shows the stability improvement for both materials.}
\label{fig:anisotropicFEM2} 
\end{figure}

\begin{figure}[h!]
\centering \includegraphics[width=1\textwidth]{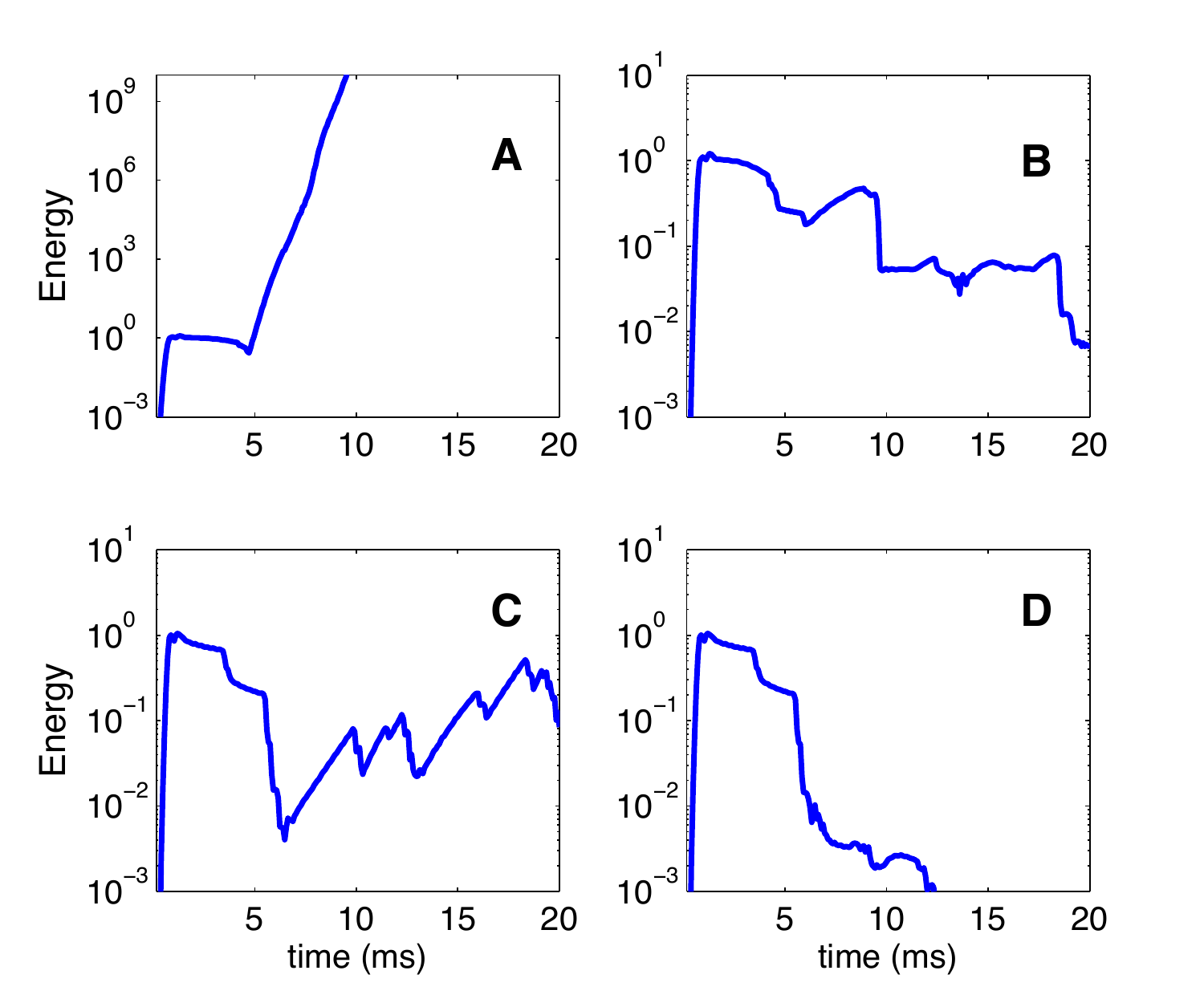}
\caption{Showing the evolution of energy in the physical domain, as represented by $\|\sqrt{v_{1}^{2}+v_{2}^{2}}\|_{\infty}$, for the same numerical experiments in figures \autoref{fig:anisotropicFEM1} and \autoref{fig:anisotropicFEM2}. (A) Material III, for the case of $\tilde{\alpha}_{j}=1$ and (B)  Material III, for the case of $\tilde{\alpha}_{1}=20,\thinspace\tilde{\alpha}_{2}=90$.(C)  Material V, for the case of $\tilde{\alpha}_{j}=1$ and (D) Material V, for the case of $\tilde{\alpha}_{1}=10,\thinspace\tilde{\alpha}_{2}=1$.}
\label{fig:energyAnisotropic} 
\end{figure}

\autoref{fig:anisotropicFEM2} shows propagation snapshots for materials III and V at the same times as in \autoref{fig:anisotropicFEM1}, but with the value $\tilde{\alpha}_{1}=1$, $\tilde{\alpha}_{2}=10$ for V, and $\tilde{\alpha}_{1}=20$, $\tilde{\alpha}_{2}=90$, and $m=n=8$ for III. Note that $\alpha_{j}$ changes from 1 to $\tilde{\alpha}_{j}$ though the PML, hence, higher order polynomial were used for high $\tilde{\alpha}_{j}$ in order to get smoother change in the PDE coefficients at the interface between the physical domain and the PML. The comparison of these two figures shows the effect of increasing the scaling parameter of the stretch function on the stability. While the instabilities disappeared for all directions in material V and in the $x_{1}$ direction for material III, some instability remained in the $x_{2}$ direction causing some energy to be reflected back to the physical domain. This is likely due to the severity of the violation of the geometric stability in the $x_{2}$ direction for this material. Nevertheless, comparing \autoref{fig:anisotropicFEM2} (C) and \autoref{fig:anisotropicFEM1} (C) (noting the use of dB scales), the use of a higher value for the scaling coefficient, $\alpha_{j}$ results in a major improvement in stability for material III. This conclusion is also evident in \autoref{fig:energyAnisotropic} that shows the manner in which the energy in the physical domain evolves in time as represented by $\left\Vert \sqrt{v_{1}^{2}+v_{2}^{2}}\right\Vert _{\infty}$ in materials III and V in both cases.

\section{Conclusions}
\label{sec:conclusions} 
Using PML approach we have addressed the problem of wave propagation in an unbounded, linear anisotropic solid in two dimensions. A time-domain second order PDE has been derived using complex coordinate stretching. An important advantage of our formulation is the small number of equations. Specifically, two second order equations along with four auxiliary equations which, to the best of knowledge, is the smallest number so far reported to describe wave propagation in solids using a time-domain PML formulation. This simplifies the problem and reduces the computational resources needed. Moreover, by reducing the formulation to a second order, use can be made of a wider variety of second order numerical schemes.

With help of the plane-wave analysis, we were able to stabilize the discrete PML problem for a wide range of otherwise unstable anisotropic media. This was achieved by increasing the value of the scaling parameter $\tilde{\alpha}_{j}$ sufficiently to move the unstable roots out of the discretely resolved range of spatial frequencies. Only two parameters stretch function was used in our formulation, while more parameters are usually used in formulations that were reported with methods to stabilize the problems. While achieving one the best reported results in stabilizing the PML problem, our method has the advantage of being simple. Discrete stability can be simply improved by increasing the value of the scaling parameter.

\section*{Acknowledgements}
The authors wish to thank Prof. Adrian Nachman and Prof. Mary Pugh of the University of Toronto Department of Mathematics for their helpful advice. RSCC is grateful to the Natural Sciences and Engineering Council (NSERC) for support under grant \#3247-2012.

\newpage
\bibliographystyle{IEEEtranN}
\bibliography{PMLRef}

\begin{thebibliography}{32}
\providecommand{\natexlab}[1]{#1}
\providecommand{\url}[1]{#1}
\csname url@samestyle\endcsname
\providecommand{\newblock}{\relax}
\providecommand{\bibinfo}[2]{#2}
\providecommand{\BIBentrySTDinterwordspacing}{\spaceskip=0pt\relax}
\providecommand{\BIBentryALTinterwordstretchfactor}{4}
\providecommand{\BIBentryALTinterwordspacing}{\spaceskip=\fontdimen2\font plus
\BIBentryALTinterwordstretchfactor\fontdimen3\font minus
  \fontdimen4\font\relax}
\providecommand{\BIBforeignlanguage}[2]{{%
\expandafter\ifx\csname l@#1\endcsname\relax
\typeout{** WARNING: IEEEtranN.bst: No hyphenation pattern has been}%
\typeout{** loaded for the language `#1'. Using the pattern for}%
\typeout{** the default language instead.}%
\else
\language=\csname l@#1\endcsname
\fi
#2}}
\providecommand{\BIBdecl}{\relax}
\BIBdecl

\bibitem[Engquist and Majda(1977)]{Engquist:1977tv}
B.~Engquist and A.~Majda, ``{Absorbing Boundary Conditions for the Numerical
  Simulation of Waves},'' \emph{Math. Comput.}, vol.~31, no. 139, pp. 629--651,
  1977.

\bibitem[B{\'e}renger(1994)]{Berenger:1994ua}
J.-P. B{\'e}renger, ``{A perfectly matched layer for the absorption of
  electromagnetic waves},'' \emph{J. Comput. Phys.}, vol. 114, no.~2, pp.
  185--200, 1994.

\bibitem[Sacks et~al.(1995)Sacks, Kingsland, and Lee]{Sacks:1995gs}
Z.~S. Sacks, D.~M. Kingsland, and R.~Lee, ``{A perfectly matched anisotropic
  absorber for use as an absorbing boundary condition},'' \emph{IEEE Trans.
  Antennas Propag.}, vol.~43, no.~12, pp. 1460--1463, 1995.

\bibitem[Roden and Gedney(1997)]{Roden:1997fo}
J.~A. Roden and S.~D. Gedney, ``{Efficient implementation of the uniaxial-based
  PML media in three-dimensional nonorthogonal coordinates with the use of the
  FDTD technique},'' \emph{Microwave Opt. Technol. Lett.}, vol.~14, no.~2, pp.
  71--75, 1997.

\bibitem[Gedney(1996)]{Gedney:1996ub}
S.~D. Gedney, ``{An anisotropic perfectly matched layer-absorbing medium for
  the truncation of FDTD lattices},'' \emph{IEEE Trans. Antennas Propag.},
  vol.~44, no.~12, pp. 1630--1639, 1996.

\bibitem[Chew and Weedon(1994)]{Chew:1994dn}
W.~C. Chew and W.~H. Weedon, ``{A 3D perfectly matched medium from modified
  maxwell's equations with stretched coordinates},'' \emph{Microwave Opt.
  Technol. Lett.}, vol.~7, no.~13, pp. 599--604, 1994.

\bibitem[Kuzuoglu and Mittra(1996)]{Kuzuoglu:1996kj}
M.~Kuzuoglu and R.~Mittra, ``{Frequency dependence of the constitutive
  parameters of causal perfectly matched anisotropic absorbers},'' \emph{IEEE
  Microw. Guided Wave Lett.}, vol.~6, no.~12, pp. 447--449, 1996.

\bibitem[Teixeira and Chew(1999)]{Teixeira:1999bt}
F.~L. Teixeira and W.~C. Chew, ``{On causality and dynamic stability of
  perfectly matched layers for FDTD simulations},'' \emph{IEEE Trans. Microwave
  Theory Tech.}, vol.~47, no.~6, pp. 775--785, 1999.

\bibitem[Meza-Fajardo and Papageorgiou(2008)]{MezaFajardo:2008dx}
K.~C. Meza-Fajardo and A.~S. Papageorgiou, ``{A Nonconvolutional, Split-Field,
  Perfectly Matched Layer for Wave Propagation in Isotropic and Anisotropic
  Elastic Media: Stability Analysis},'' \emph{Bull. Seismol. Soc. Am.},
  vol.~98, no.~4, pp. 1811--1836, 2008.

\bibitem[B{\'e}renger(2002)]{Berenger:2002bq}
J.-P. B{\'e}renger, ``{Application of the CFS PML to the absorption of
  evanescent waves in waveguides},'' \emph{IEEE J. Sel. Areas Commun.},
  vol.~12, no.~6, pp. 218--220, 2002.

\bibitem[B{\'e}cache et~al.(2004)B{\'e}cache, Petropoulos, and
  Gedney]{Becache:2004fz}
E.~B{\'e}cache, P.~G. Petropoulos, and S.~D. Gedney, ``{On the long-time
  behavior of unsplit perfectly matched layers},'' \emph{IEEE Trans. Antennas
  Propag.}, vol.~52, no.~5, pp. 1335--1342, 2004.

\bibitem[Duru and Kreiss(2012)]{Duru:2012va}
K.~Duru and G.~Kreiss, ``{A well-posed and discretely stable perfectly matched
  layer for elastic wave equations in second order formulation},''
  \emph{Commun. Comput. Phys.}, vol.~11, no.~5, pp. 1643--1672, 2012.

\bibitem[Appel{\"o} and Kreiss(2006)]{Appelo:2006wz}
D.~Appel{\"o} and G.~Kreiss, ``{A new absorbing layer for elastic waves},''
  \emph{J. Comput. Phys.}, vol. 215, no.~2, pp. 642--660, 2006.

\bibitem[Collino and Tsogka(2001)]{Collino:2001vt}
F.~Collino and C.~Tsogka, ``{Application of the perfectly matched absorbing
  layer model to the linear elastodynamic problem in anisotropic heterogeneous
  media},'' \emph{Geophysics}, vol.~66, no.~1, pp. 294--307, 2001.

\bibitem[Hastings et~al.(1996)Hastings, Schneider, and
  Broschat]{Hastings:1996um}
F.~Hastings, J.~B. Schneider, and S.~L. Broschat, ``{Application of the
  perfectly matched layer (PML) absorbing boundary condition to elastic wave
  propagation},'' \emph{J. Acoust. Soc. Am.}, vol. 100, no.~5, pp. 3061--3069,
  1996.

\bibitem[Drossaert and Giannopoulos(2007{\natexlab{a}})]{Drossaert:2007fi}
F.~H. Drossaert and A.~Giannopoulos, ``{Complex frequency shifted convolution
  PML for FDTD modelling of elastic waves},'' \emph{Wave Motion}, vol.~44, no.
  7-8, pp. 593--604, 2007.

\bibitem[Chew and Liu(1996)]{Chew:1996wk}
W.~C. Chew and Q.-H. Liu, ``{Perfectly matched layers for elastodynamics: A new
  absorbing boundary condition},'' \emph{J. Comput. Acoust.}, vol.~4, no.~4,
  pp. 341--359, 1996.

\bibitem[Kucukcoban and Kallivokas(2011)]{Kucukcoban:2011tn}
S.~Kucukcoban and L.~F. Kallivokas, ``{Mixed perfectly-matched-layers for
  direct transient analysis in 2D elastic heterogeneous media},'' \emph{Comput.
  Meth. Appl. Mech. Eng.}, vol. 200, no. 1-4, pp. 57--76, 2011.

\bibitem[Appel{\"o} et~al.(2006)Appel{\"o}, Hagstrom, and
  Kreiss]{Appelo:2006vv}
D.~Appel{\"o}, T.~Hagstrom, and G.~Kreiss, ``{Perfectly Matched Layers for
  Hyperbolic Systems: General Formulation, Well-posedness, and Stability},''
  \emph{SIAM J. Appl. Math.}, vol.~67, no.~1, pp. 1--23, 2006.

\bibitem[Roden and Gedney(2000)]{Roden:2000cn}
J.~A. Roden and S.~D. Gedney, ``{Convolution PML (CPML): An efficient FDTD
  implementation of the CFS-PML for arbitrary media},'' \emph{Microwave Opt.
  Technol. Lett.}, vol.~27, no.~5, pp. 334--339, 2000.

\bibitem[Loh et~al.(2009)Loh, Oskooi, Ibanescu, Skorobogatiy, and
  Johnson]{Loh:2009uc}
P.~R. Loh, A.~F. Oskooi, M.~Ibanescu, M.~Skorobogatiy, and S.~G. Johnson,
  ``{Fundamental relation between phase and group velocity, and application to
  the failure of perfectly matched layers in backward-wave structures},''
  \emph{Phys. Rev. E}, vol.~79, no.~6, 2009.

\bibitem[Kreiss and Duru(2013)]{Kreiss:2013tb}
G.~Kreiss and K.~Duru, ``{Discrete stability of perfectly matched layers for
  anisotropic wave equations in first and second order formulation},''
  \emph{BIT Numer. Math.}, vol.~53, no.~3, pp. 641--663, Mar. 2013.

\bibitem[Komatitsch and Martin(2007)]{Komatitsch:2007bz}
D.~Komatitsch and R.~Martin, ``{An unsplit convolutional perfectly matched
  layer improved at grazing incidence for the seismic wave equation},''
  \emph{Geophysics}, vol.~72, no.~5, p. SM155, 2007.

\bibitem[Beltzer(1988)]{Beltzer:1988ut}
A.~I. Beltzer, \emph{{Acoustics of solids}}.\hskip 1em plus 0.5em minus
  0.4em\relax New York: Springer-Verlag, 1988.

\bibitem[B{\'e}cache et~al.(2003)B{\'e}cache, Fauqueux, and
  Joly]{Becache:2003ug}
E.~B{\'e}cache, S.~Fauqueux, and P.~Joly, ``{Stability of perfectly matched
  layers, group velocities and anisotropic waves},'' \emph{J. Comput. Phys.},
  vol. 188, no.~2, pp. 399--433, 2003.

\bibitem[Johnson(2008)]{Johnson:2008wt}
S.~G. Johnson. (2008) {Notes on Perfectly Matched Layers ( PMLs), MIT Open
  Course Ware, }.

\bibitem[Teixeira and Chew(2000)]{Teixeira:2000vj}
F.~L. Teixeira and W.~C. Chew, ``{Complex space approach to perfectly matched
  layers: a review and some new developments},'' \emph{Int. J. Numer. Modell.
  Electron. Networks Devices Fields}, vol.~13, no.~5, pp. 441--455, 2000.

\bibitem[Zhang and Shen(2010)]{Zhang:2010vp}
W.~Zhang and Y.~Shen, ``{Unsplit complex frequency-shifted PML implementation
  using auxiliary differential equations for seismic wave modeling},''
  \emph{Geophysics}, vol.~75, no.~4, pp. T141--T154, 2010.

\bibitem[Drossaert and Giannopoulos(2007{\natexlab{b}})]{Drossaert:2007bv}
F.~H. Drossaert and A.~Giannopoulos, ``{A nonsplit complex frequency-shifted
  PML based on recursive integration for FDTD modeling of elastic waves},''
  \emph{Geophysics}, vol.~72, no.~2, p.~T9, 2007.

\bibitem[Petropoulos(2000)]{Petropoulos:2000wh}
P.~G. Petropoulos, ``{Reflectionless Sponge Layers as Absorbing Boundary
  Conditions for the Numerical Solution of Maxwell Equations in Rectangular,
  Cylindrical, and Spherical Coordinates},'' \emph{SIAM J. Appl. Math.},
  vol.~60, no.~3, pp. 1037--1058, 2000.

\bibitem[Daros(2007)]{Daros:2007de}
C.~H. Daros, ``{Material Stability Conditions for a Class of Inhomogeneous
  Anisotropic Media},'' \emph{Math. Mech. Solids}, vol.~14, no.~4, pp.
  377--389, 2007.

\bibitem[Li and Matar(2010)]{Li:2010tu}
Y.~Li and O.~B. Matar, ``{Convolutional perfectly matched layer for elastic
  second-order wave equation},'' \emph{J. Acoust. Soc. Am.}, vol. 127, no.~3,
  pp. 1318--1327, 2010.

\end{thebibliography}

\end{document}